\documentstyle{mn}

\newif\ifAMStwofonts
\def \SAIT #1 #2 {{\em Mem.\ Soc.\ Astron.\ It.\/} {\bf #1}, #2}
\def \MESS #1 #2 {{\em The Messenger\/} {\bf #1}, #2}
\def \ASTRNACH #1 #2 {{\em Astron. Nach.\/} {\bf #1}, #2}
\def \AAP #1 #2 {{\em Astron. Astrophys.\/} {\bf #1}, #2}
\def \AAL #1 #2 {{\em Astron. Astrophys. Lett.\/} {\bf #1}, L#2}
\def \AAR #1 #2 {{\em Astron. Astrophys. Rev.\/} {\bf #1}, #2}
\def \AAS #1 #2 {{\em Astron. Astrophys. Suppl. Ser.\/} {\bf #1}, #2}
\def \AJ #1 #2 {{\em Astron. J.\/} {\bf #1}, #2}
\def \ANNREV #1 #2 {{\em Ann. Rev. Astron. Astrophys.\/} {\bf #1}, #2}
\def \APJ #1 #2 {{\em Astrophys. J.\/} {\bf #1}, #2}
\def \APJL #1 #2 {{\em Astrophys. J. Lett.\/} {\bf #1}, L#2}
\def \APJS #1 #2 {{\em Astrophys. J. Suppl.\/} {\bf #1}, #2}
\def \APSS #1 #2 {{\em Astrophys. Space Sci.\/} {\bf #1}, #2}
\def \ASR #1 #2 {{\em Adv. Space Res.\/} {\bf #1}, #2}
\def \BAIC #1 #2 {{\em Bull. Astron. Inst. Czechosl.\/} {\bf #1}, #2}
\def \JSQRT #1 #2 {{\em J. Quant. Spectrosc. Radiat. Transfer\/} {\bf #1}, #2}
\def \MN #1 #2 {{\em Mon. Not. R. Astr. Soc.\/} {\bf #1}, #2}
\def \MEM #1 #2 {{\em Mem. R. Astr. Soc.\/} {\bf #1}, #2}
\def \PLR #1 #2 {{\em Phys. Lett. Rev.\/} {\bf #1}, #2}
\def \PASJ #1 #2 {{\em Publ. Astron. Soc. Japan\/} {\bf #1}, #2}
\def \PASP #1 #2 {{\em Publ. Astr. Soc. Pacific\/} {\bf #1}, #2}
\def \NAT #1 #2 {{\em Nature\/} {\bf #1}, #2}
\def\strutdepth{\dp\strutbox}
\def\marginalstar{\strut\vadjust{\kern-\strutdepth\specialstar}}
\def\specialstar{\vtop to \strutdepth{\baselineskip\strutdepth\vss\llap{$\bigtriangleup\mskip-12mu\bigtriangledown$ }\null}}
\def \nm {N_{\mathrm{m}}}
\def \nm0 {N_{\mathrm{m,0}}}
\def \na {N_{\mathrm{a}}}
\def \na0 {N_{\mathrm{a,0}}}

\def \BGE {\begin{equation}}
\def \EDE {\end{equation}}

\ifoldfss
  \ifCUPmtlplainloaded \else
    \NewTextAlphabet{textbfit} {cmbxti10} {}
    \NewTextAlphabet{textbfss} {cmssbx10} {}
    \NewMathAlphabet{mathbfit} {cmbxti10} {} % for math mode
    \NewMathAlphabet{mathbfss} {cmssbx10} {} %  "   "    "
  \fi
  \ifAMStwofonts
    \ifCUPmtlplainloaded \else
      \NewSymbolFont{upmath} {eurm10}
      \NewSymbolFont{AMSa} {msam10}
      \NewMathSymbol{\upi}     {0}{upmath}{19}
      \NewMathSymbol{\umu}     {0}{upmath}{16}
      \NewMathSymbol{\upartial}{0}{upmath}{40}
      \NewMathSymbol{\leqslant}{3}{AMSa}{36}
      \NewMathSymbol{\geqslant}{3}{AMSa}{3E}

      \let\geq=\geqslant 
    \fi
  \fi
\fi % End of OFSS

\ifnfssone
  \newmathalphabet{\mathit}
  \addtoversion{normal}{\mathit}{cmr}{m}{it}
  \addtoversion{bold}{\mathit}{cmr}{bx}{it}
  \newmathalphabet{\mathbfit} % math mode version of \textbfit{..}
  \addtoversion{normal}{\mathbfit}{cmr}{bx}{it}
  \addtoversion{bold}{\mathbfit}{cmr}{bx}{it}
  \newmathalphabet{\mathbfss} % math mode version of \textbfss{..}
  \addtoversion{normal}{\mathbfss}{cmss}{bx}{n}
  \addtoversion{bold}{\mathbfss}{cmss}{bx}{n}
  \ifAMStwofonts
    \ifCUPmtlplainloaded \else
      \UseAMStwoboldmath
      \makeatletter
      \new@mathgroup\upmath@group
      \define@mathgroup\mv@normal\upmath@group{eur}{m}{n}
      \define@mathgroup\mv@bold\upmath@group{eur}{b}{n}
      \edef\UPM{\hexnumber\upmath@group}
      \new@mathgroup\amsa@group
      \define@mathgroup\mv@normal\amsa@group{msa}{m}{n}
      \define@mathgroup\mv@bold\amsa@group{msa}{m}{n}
      \edef\AMSa{\hexnumber\amsa@group}
      \makeatother
      \mathchardef\upi="0\UPM19
      \mathchardef\umu="0\UPM16
      \mathchardef\upartial="0\UPM40
      \mathchardef\leqslant="3\AMSa36
      \mathchardef\geqslant="3\AMSa3E

      \let\geq=\geqslant 
    \fi
  \fi
\fi % End of NFSS release 1

\ifnfsstwo
  \DeclareMathAlphabet{\mathbfit}{OT1}{cmr}{bx}{it}
  \SetMathAlphabet\mathbfit{bold}{OT1}{cmr}{bx}{it}
  \DeclareMathAlphabet{\mathbfss}{OT1}{cmss}{bx}{n}
  \SetMathAlphabet\mathbfss{bold}{OT1}{cmss}{bx}{n}
  \ifAMStwofonts
    \ifCUPmtlplainloaded \else
      \DeclareSymbolFont{UPM}{U}{eur}{m}{n}
      \SetSymbolFont{UPM}{bold}{U}{eur}{b}{n}
      \DeclareSymbolFont{AMSa}{U}{msa}{m}{n}
      \DeclareMathSymbol{\upi}{0}{UPM}{"19}
      \DeclareMathSymbol{\umu}{0}{UPM}{"16}
      \DeclareMathSymbol{\upartial}{0}{UPM}{"40}
      \DeclareMathSymbol{\leqslant}{3}{AMSa}{"36}
      \DeclareMathSymbol{\geqslant}{3}{AMSa}{"3E}

      \let\geq=\geqslant 
    \fi
  \fi
\fi % End of NFSS release 2

\ifCUPmtlplainloaded \else
  \ifAMStwofonts \else % If no AMS fonts
    \def\upi{\pi}
    \def\umu{\mu}
    \def\upartial{\partial}
  \fi
\fi
   \title[The first world atlas of the artificial night sky brightness]{The first world atlas of the artificial night sky brightness}

   \author[P. Cinzano,        
          F. Falchi and 
          C. D. Elvidge 
           ]{P. Cinzano$^{1}$\thanks{E-mail: cinzano@pd.astro.it, cinzano@lightpollution.it}\thanks{also at the Istituto di Scienza e Tecnologia dell'Inquinamento Luminoso (ISTIL), Thiene, Italy},        
          F. Falchi$^{1 \dag}$,         and
          C. D. Elvidge$^2$ \\
$^{1}$ Dipartimento di Astronomia, Universit\`a di Padova,
vicolo dell'Osservatorio 5,  I-35122 Padova, Italy\\
$^2$ Office of the Director, NOAA National Geophysical Data Center, 325 Broadway, Boulder CO 80303}

\date{Accepted 1 August 2001.
      Received 24 July 2001;
      in original form 18 December 2000}

\pagerange{\pageref{firstpage}--\pageref{lastpage}}
\pubyear{2001}

\input epsf.sty
\begin{document}

\maketitle

\label{firstpage}

\begin{abstract}

We present the first World Atlas of the zenith artificial night sky brightness at sea level. Based on radiance calibrated high resolution DMSP satellite data and on accurate modelling of light propagation in the atmosphere, it provides a nearly global picture of how mankind is proceeding to envelope itself in a luminous fog.
Comparing the Atlas with the U.S. Department of Energy (DOE) population density database we determined the fraction of population who are living under a sky of given brightness. About two thirds of the World population and 99\% of the population in US (excluding Alaska and Hawaii) and EU live in areas where the night sky is above the threshold set for polluted status. Assuming average eye functionality, about one fifth of the World population, more than two thirds of the US population and more than one half of the EU population have already lost naked eye visibility of the Milky Way. Finally, about one tenth of the World population, more than 40\% of the US population and one sixth of the EU population no longer view the heavens with the eye adapted to night vision because the sky brightness.

\end{abstract}

\begin{keywords}
atmospheric effects
               -- site testing
               -- scattering -- 
                 light pollution
\end{keywords}

%
%________________________________________________________________

\section{Introduction}

One of the most rapidly increasing alterations to the natural environment is the alteration of the ambient light levels in the night environment produced by man-made light. The study of global change must take into account this phenomenon called light pollution. Reported adverse effects of light pollution involve the animal kingdom, the vegetable kingdom and mankind (see e.g. Cinzano 1994 for a reference list). Moreover, the growth of the night sky brightness associated with light pollution produces a loss of perception of the Universe where we live (see e.g. Crawford 1991; Kovalevsky 1992; McNally 1994; Isobe \& Hirayama 1998; Cinzano 2000d; Cohen \& Sullivan 2001).  This could have unintended impacts on the future of our society. In fact  the night sky, which constitutes the panorama of the surrounding Universe, has always had a strong influence on human thought and culture, from philosophy to religion, from art to literature and science. 

Interest in light pollution has been growing in many fields of science, extending from the traditional field of astronomy, to atmospheric physics, environmental sciences, natural sciences and even human sciences. The full extent and implications of the problem have not been addressed to date due to the fact that there have been no global-scale data on the distribution and magnitude of artificial sky brightness.

The zenith artificial night sky brightness at sea level is a useful indicator of the effects of light pollution on the night sky and the atmospheric content of artificial light. Sea level maps of it, being free of elevation effects, are useful for comparing pollution levels across large territories, for recognizing the most polluted areas or more polluting cities and for identifying dark areas (Cinzano et al. 2000a, hereafter Paper 1).  Even if the capability to perceive the Universe is better shown by specific maps of stellar visibility, which account for altitude and atmospheric extinction (Cinzano et al. 2000b, hereafter Paper 2), maps of the zenith artificial sky brightness at sea level provide a reasonable statistical evaluation of the visibility of the Milky Way and a comparison with typical natural brightness levels.  The sea level product is also a reasonable starting point in the global study of light pollution given that population numbers are concentrated at low altitudes.

To date no global, quantitative and accurate depiction of the artificial brightness of the night sky has been available to the scientific community and governments. Ground based measurements of sky brightness are available only for a limited number of sites, mainly astronomical observatories, and are spread over many different years.  The paucity of ground based observations makes it impossible to construct global maps from this source. 

One approach to modelling the spatial distribution of artificial night sky brightness is to predict it based on population density, since areas with high population usually produce higher levels of light pollution and, consequently, a high artificial luminosity of the night sky (sky glow). However (i) the apparent proportionality between population and sky glow breaks down going from large scales to smaller scales and looking in more detail, owing to the atmospheric propagation of light pollution large distances from the sources, (ii)  the upward light emission is not always proportional to the population (e.g. due to differences in development and lighting practices), (iii) some polluting sources are not represented in population data (e.g. industrial sites and gas flares) and (iv) population census data are not collected using uniform techniques, timetables, or administrative reporting units around the world. 

As an alternative, we have used a global map of top of atmosphere radiances from manmade light sources produce using data from the U.S. Air Force Defense Meteorological Satellite Program (DMSP)Operational Linescan System (OLS) to model artificial sky brightness.  From 1972-92 only film data were available from the DMSP-OLS.  Sullivan (1989, 1991) was successful in producing a global map of light sources using film data, but this product did not distinguish between the persistent light sources of cities and the ephemeral lights of events such as fire.  In the mid-1990s Elvidge et al. (1997a,b,c) produced a global cloud-free composite of lights using a time series of DMSP nighttime observations, identifying the locations of persistent light sources.  This potential use of these "stable lights" for light pollution studies was noted by Isobe and Hamamura (1998). More recently a radiance calibrated global map of manmade light sources has been produced using DMSP-OLS data collected at reduced gain settings (Elvidge et al. 1999). With both the location and top of atmosphere radiances mapped, the stage was set to model artificial sky brightness across the world's surface.

The first exploration of these data for predicting artificial sky brightness were made by applying simple light pollution propagation laws to the satellite data (Falchi 1998; Falchi \& Cinzano 2000).  Subsequently we introduced a method to map the artificial sky brightness (Paper 1) and naked-eye star visibility (Paper 2) across large territories, computing the propagation of light inside the atmosphere using the detailed Garstang Models (Garstang 1984, 1986, 1898a, 1989b, 1991, 2000; see also Cinzano 2000a,b).  Here we present the first World Atlas of the zenith artificial night sky brightness at sea level. It has been obtained by applying the method discussed in Paper 1 to global high resolution radiance calibrated DMSP satellite data.
In sec. \ref{s2} we summarize the outline of the method, in sec. \ref{s3} we present the Atlas and a comparison with Earth-based measurements, in sec. \ref{s4b} we present statistical results and tables based on a comparison with the Landscan 2000 DOE population density database (Dobson et al. 2000) and in section \ref{s4} we draw our conclusions.

\section{Outlines of the method}
\label{s2}

Here we summarize the methods used to produce the World Atlas. We refer the readers to Paper 1 and Paper 2 for a detailed discussion.

High resolution upward flux data have been calculated from radiances observed by the Operational Linescan System (OLS) carried by the DMSP satellites.  The OLS is an oscillating scan radiometer with low-light visible and thermal infrared (TIR) imaging capabilities (Lieske 1981). At night the OLS uses a Photo Multiplier Tube (PMT), attached to a 20 cm reflector telescope, to intensify the visible band signals. It has a broad spectral response from 440 to 940 nm  with highest sensitivity in the 500 to 650 nm region, covering the range for primary emissions from the most widely used lamps for external lighting:  Mercury Vapour (545 nm and 575 nm), High Pressure Sodium (from 540 nm to 630 nm) and Low Pressure Sodium (589 nm).  We used a global map of radiances produces using 28 nights of data collected in 1996-97 at reduced gain levels, to avoid saturation in urban centers.  The global map is a "cloud-free" composite, meaning that only cloud-free observations were used.  The map reports the average radiance observed from the set of cloud-free observations.   Ephemeral lights produced by fires and random noise events were removed by deleting lights which occurred in the same place less than three times. Calibrated upward fluxes per unit solid angle toward the satellite have been obtained from radiance data based on a pre-flight irradiance calibration of the OLS photomultiplier tube (PMT). The calibration was tested with Earth-based measurements in Paper 1. The upward flux per unit solid angle in other directions was estimated based on an average normalized emission function, in agreement with a study of the upward flux per unit solid angle per inhabitant of a large number of cities at different distances from the satellite nadir.

The propagation of light pollution is computed with the Garstang modelling techniques taking into account Rayleigh scattering by molecules, Mie scattering by aerosols, atmospheric extinction along light paths and Earth curvature. We neglected third and higher order scattering which can be significant only for optical thicknesses higher than ours. We associated the predictions with well-defined parameters related to the aerosol content, so the atmospheric conditions, which predictions involve, are well known. Atmospheric conditions are variable and a careful evaluation of the "typical" atmospheric condition in the local "typical" clear night of each area is quite difficult, even due to the difficulty to define it, so we used the same atmospheric model everywhere, corresponding to a standard clean atmosphere (Garstang 1986, 1989; Paper 1; Paper 2). This also avoids confusion between effects due to light pollution and effects due to geographic gradients of atmospheric conditions in "typical" nights. 
Being more interested in understanding and comparing light pollution distributions rather than in predicting the effective sky brightness for observational purposes, we computed the artificial sky brightness at sea level, in order to avoid the introduction of altitude effects into our maps. 
Readers should consider these differences when interpreting the Atlas results and the related statistics. 

\section{Results}
\label{s3}

The World Atlas of the Sea Level Artificial Night Sky Brightness has been computed for the photometric astronomical V band, at the zenith, for a clean atmosphere with an aerosol clarity coefficient K=1, where K is a coefficient which measures the aerosol content of the atmosphere (Garstang 1986), corresponding to a vertical extinction $\Delta m=$0.33 mag in the V band,  a horizontal visibility $\Delta x=$26 km and an optical depth $\tau=$0.3.  The maps of each continent are shown in figure 1 to figure 8 in latitude/longitude projection.  The original high resolution maps of the World Atlas are downloadable as zipped {\sc TIFF }files from the World Wide Web site http://www.lightpollution.it/dmsp/.  They have been obtained with a mosaic of the original $30''\times30''$ pixel size maps.
Each map level is three times larger than the previous one. The map levels correspond to the artificial sky brightnesses (between brackets the respective colours) in V $\mathrm{ph\, cm^{-2} s^{-1} sr^{-1}}$: $9.47\;10^{6}-2.84\;10^{7}$  (blue), $2.84\;10^{7}-8.61\;10^{7}$ (green), $8.61\;10^{7}-2.58\;10^{8}$  (yellow), $2.58\;10^{8}-7.75\;10^{8}$ (orange), $7.75\;10^{8}-2.32\;10^{9}$  (red), $>$$2.32\;10^{9}$ (white),
or in $\mu \mathrm{cd/m^{2}}$: 27.7-83.2 (blue), 83.2-252 (green), 252-756 (yellow), 756-2268 (orange), 2268-6804 (red), $>$6804 (white)(based on the conversion in Garstang 1986, 1989). 
For the dark-grey level see below.
The map levels can be expressed more intuitively as ratios between the artificial sky brightness and the reference natural sky brightness. The natural night sky brightness depends on the geographical position, the solar activity, the time from the sunset and the sky area observed (see e.g. paper 2), so we referred the levels in our maps to an average sky brightness below the atmosphere of $b_{n}=8.61\;10^{7}$ V $\mathrm{ph\, cm^{-2} s^{-1} sr^{-1}}$, corresponding approximately to 21.6 V $\mathrm{mag/arcsec^{2}}$ or 252 $\mu \mathrm{cd/m^{2}}$ (Garstang 1986).  In this case the map levels became: 0.11-0.33 (blue), 0.33-1 (green), 1-3 (yellow), 3-9 (orange), 9-27 (red), $>$27 (white).  Country boundaries are approximate.
In order to show how far the light pollution propagates from sources, we coloured  in dark-grey areas where the artificial sky brightness is greater than 1\% of the reference natural brightness (i.e. greater than $8.61\;10^{5}$ V $\mathrm{ph\, cm^{-2} s^{-1} sr^{-1}}$ or 2.5 $\mu \mathrm{cd/m^{2}}$).
In these areas the night sky can be considered unpolluted at the zenith but at lower elevations pollution might be not negligible and uncontrolled growth of light pollution will endanger even the zenith sky. This level must be considered only an indication because small differences in atmospheric conditions can produce large differences where the gradient of artificial brightness is small. 

The resolution of the ATLAS does not correspond directly to the DMSP-OLS pixel size. The effective instantaneous field of view (EIFOV) of OLS-PMT is larger than the pixel-to-pixel ground sample distance maintained by the along-track OLS sinusoidal scan and the electronic sampling of the signal from the individual scan lines.  Moreover the original data have been "smoothed" by on-board averaging of  5 by 5 pixel blocks, yielding a ground sample distance of 2.8 km.  During geolocation the OLS pixel values are used to fill 30 arc second grids, which are composited to generate the global 30 arc second grid. However, since the sky brightness is frequently produced by the sum of many contributions from distant sources, the lower resolution of the upward flux data do not play a role and the map resolution mainly corresponds to the 30 arc second grid cell size which at equator is 0.927 km. 

The satellite data also record the offshore lights where oil and gas production is active (visible e.g. in the North Sea, Chinese Sea and Arabic Gulf), other natural gas flares (visible e.g. in Nigeria) and the fishing fleets (visible e.g. near the coast of Argentina, in the Japan Sea and near Malacca). Their upward emission functions likely differ from the average emission function of the urban night-time lighting so that the predictions of their effects have some uncertainty. The presence of snow could also add some uncertainty (see Paper 1). For this reasons we neglected territories near the poles.

The differences between the levels for Europe in figure 3, based on the pre-flight OLS-PMT radiance calibration and referring to 1996-1997, and in figures 11 and 12 of Paper 1, based on  calibration with Earth-based measurements and referring to 1998-1999, agree with the yearly growth of light pollution measured in Europe (see e.g. Cinzano 2000c) but they cannot be considered significant because they are within the uncertainties of the method.

A comparison between map predictions and Earth-based sky brightness measurements is presented in figure 7. 
The left panel shows map predictions versus artificial night sky brightness measurements at the bottom of the atmosphere taken in clean or photometric nights in the V band for Europe (filled squares), North America (open triangles), South America (open rhombi), Africa (filled triangles), Asia (filled circle) (Catanzaro \& Catalano 2000; Della Prugna 2000; Falchi 1998; Favero et al. 2000; Massey \& Foltz 2000; Nawar et al. 1998; Nawar et al. 1998; Piersimoni et al. 2000; Poretti \& Scardia 2000; Zitelli 2000). All of them have been taken in 1996-1997 except those for Europe which have been taken in 1998-1999 and rescaled to 1996-1997 by subtracting 20\% in order to approximately account for the growth of light pollution in two years. Errorbars account for measurement errors and for an uncertainty of about 0.1 mag $\mathrm{arcsec}^{2}$ in the subtracted natural sky brightness which is non-negligible in dark sites. These are smaller than the effects of fluctuations in atmospheric conditions. The right panel shows  map predictions versus photographic measurements taken in Japan in the period 1987-1991 with variable atmospheric aerosol content (Kosai et al. 1992). They are calibrated to the top of the atmosphere and averaged for each site neglecting those where less than five measurements were taken. The large errorbars show the effects of  changes in the atmospheric aerosol content and in the extinction of the light of the comparison star. The dashed line shows the linear regression.
A worldwide project of the International Dark-Sky Association (IDA) is collecting a large number of accurate CCD measurements of sky brightness together with the aerosol content, which could be valuable for testing future improvements in the modelling of artificial sky brightness (Cinzano \& Falchi 2000).

\section{Statistics}
\label{s4b}

We compared our Atlas with the Landscan 2000 DOE global population density database (Dobson et al. 2000) which has the same 30 arc second grid cell size as our Atlas. We checked the spatial match of our Atlas against the Landscan data by visual inspection of the superimposition of the two datasets. We extracted statistics for each individual countries, for the European Union and for the World, tallying the percent population who on standard clear atmosphere nights are living inside each level of our Atlas. Additionally we tallied the percentage of population living under a sky brightness greater than several other sky brightness conditions, as described below. Table 1 shows the percentage of population who are living under a sky brightness greater than each level of our Atlas in standard clean nights, i.e. the ratios between the artificial sky brightness and the reference natural sky brightness are greater than  0.11 (column 1), 0.33 (column 2), 1 (column 3), 3 (column 4), 9 (column 5), 27 (column 6). The table also shows the fraction of population who in standard clean nights are living under a sky brightness greater than some typical sky brightnesses: the threshold $b_{p}$ to consider the night sky polluted (i.e. when the artificial sky brightness is greater than 10\% of the natural night sky brightness above 45 degrees of elevation (Smith 1979)) (column 7), the sky brightness $b_{fq}$ measured with a first quarter moon in the best astronomical sites  (e.g. Walker 1987)(column 8), the sky brightness $b_{m}$ in the considered location with a first quarter moon at 15 degrees elevation  (based on Krisciunas \& Schaefer 1991) and zero light pollution (column 9), the sky brightness $b_{fm}$ measured close to full moon in the best astronomical sites  (e.g. Walker 1987)(column 10) which is not much larger than the typical zenith brightness at nautical twilight (Schaefer 1993), the threshold of visibility of the Milky Way for average eye capability $b_{mw}$ (column 11), the eye's night vision threshold $b_{e}$ (Garstang 1986; see also Schaefer 1993)(column 12). Table 2 resumes their numerical values.

To produce the Landscan, DOE collected the best available census data for each country and calculated a probability coefficient for the population density of each 30 arc second grid cell.  The probability coefficient is based on slope, proximity to roads, land cover, nighttime lights, and an urban density factor (Dobson et al. 2000).  The probability coefficients are used to perform a spatial allocation of the population for all the grid cells covering a census reporting unit (usually province).  Therefore the resulting population distribution represents an ambient population which integrates diurnal movements and collective travel habits rather than the residential population at nighttime.  Readers must be aware that these percentages should be considered as estimates due to the proceeding discussion on the Landscan data characteristics, the minor altitude effects on the artificial sky brightness levels and departures in the angular distribution of light from sources from the assumed average normalized emission function.

We also determined the surface area corresponding to each level of our Atlas. Table 3 shows the fraction per cent of the surface area of the individual World countries,  the European Union and  the World, where the sky brightness is greater than each level of our Atlas in standard clean nights, i.e. the ratios between the artificial sky brightness and the reference natural sky brightness are greater than  0.11 (column 1), 0.33 (column 2), 1 (column 3), 3 (column 4), 9 (column 5), 27 (column 6). 

Figure \ref{f9} shows in white the World's area covered by our Atlas where 98\% of the World population lives. Our data refer to 1996-1997, so the artificial night sky brightness today is likely increased. 

\section{Conclusions}
\label{s4}

The Atlas  reveals that light pollution of the night sky is not confined, as commonly believed, to developed countries, but rather appears to be a global-scale problem affecting nearly every country of the World. The problem is more severe in the US, Europe and Japan, as expected. However the night sky appears more seriously endangered than commonly believed. 

The population percentages presented in Tables 1 and 3 speak for themselves, indicating that large numbers of people in many countries have had their vision of the night sky severely degraded. Our Atlas refers to 1996-1997, so the situation today is undoubtably worse. 
We found that more than 99\% of the US and EU population, and about two thirds of the World population live in areas where the night sky is above the threshold considered polluted (i.e. the artificial sky brightness is greater than 10\% of the natural night sky brightness above 45 degrees of elevation (Smith 1979)). In the areas where 97\% of the US population, 96\% of the EU population and half of the World population live, the night sky in standard clean atmospheric conditions is brighter than has been measured with a first quarter moon in the best astronomical sites (e.g. Walker 1987). 93\% of the US population, 90\% of the EU population and about 40\% of the World population live under a zenith night sky which is brighter than they would have in the same location with a first quarter moon at 15 degrees elevation (based on Krisciunas \& Schaefer 1991) and zero light pollution. So they effectively live in perennial moonlight. They rarely realize it because they still experience the sky to be brighter under a full moon than under new moon conditions.  We also found that for about 80\% of the US population, two thirds of the EU population and more than one fourth of the World population the sky brightness is even greater than that measured close to full moon in the best astronomical sites (e.g. Walker 1987). "Night" never really comes for them because this sky brightness is approximately equal to the typical zenith brightness at nautical twilight (Schaefer 1993). Assuming average eye functionality, more than two thirds of the US population, about half of the EU population and one fifth of the World population have already lost the possibility to see the Milky Way, the galaxy where we live. Finally, approximately 40\% of the US population, one sixth of the EU population and one tenth of the World population cannot even look at the heavens with the eye adapted to night vision because its brightness is above the night vision threshold (Garstang 1986; see also Schaefer 1993). Preliminary data on moonlight without the moon was presented by Cinzano et al. (2001).

We noticed that Venice is the only city in Italy with more than 250000 inhabitants from which an average observer has the possibility to view the Milky Way from the city center on a clear night in 1996-97.   Even though the Venice's historic centre (pop. 68000) is imbedded in the strong sky glow produced by the terra firma part of the city (Mestre, pop. 189000), its average artificial sky brightness is still lower than in cities with 80.000 inhabitants in the nearby Veneto plane. This is due mainly to the unique low intensity romantic lighting of this city, which deserves to be preserved. 

Many areas which were believed to be unpolluted because they appear completely dark in night-time satellite images, on the contrary show in the Atlas non-negligible artificial brightness levels, due to the outward propagation of light pollution. In a number of cases the sky of a country appears polluted by sources in a neighbouring country: this could open a new chapter of international jurisprudence. Astronomical observatories known for their negligible zenith artificial sky brightness appear to lie near or inside the 1\% level: this means that without undertaking a serious control of light pollution in liable areas they risk in less than 20 years seeing their sky quality degraded. Site testing for next generation telescopes will require an accurate study of the long-term growth of the artificial night sky brightness in order to assure dark sky conditions for an adequate number of years after their installation. Serious control both of lighting installations and of new urbanizations or developments would be necessary for a large area surrounding the site (possibly even 250 km in radius).

We are working to the preparation of a forecoming Atlas giving the growth rates of light pollution, the growth rates of night sky brightness, the emission functions of the sources (Paper 1) and the ratio of the upward light flux versus population per unit area.

The International Dark-Sky Association (http://www.darksky.org) is supporting worldwide the legislative effort carried on in many countries to limit light pollution, in order to protect astronomical observatories, amateurs observatories, the citizens' perception of the universe, the environment and to save energy, money and resources. Commission 50 of the International Astronomical Union (``The protection of existing and potential astronomical sites'') is working actively to preserve the astronomical sky, now with a specific Working Group (``Controlling light pollution'') born after the  UN-IAU Special Environmental Symposium "Preserving the Astronomical Sky" held in the Vienna United Nations Organization's Centre in the summer of 1999 (Cohen \& Sullivan 2000). 

\section*{Acknowledgments}
We are grateful to Roy Garstang of JILA-University of Colorado for his friendly kindness in reading and commenting on this paper, for his helpful suggestions and for interesting discussions. We acknowledge the unknown referee for the stimulus to extend this work with statistical tables. PC acknowledges the Istituto di Scienza e Tecnologia dell'Inquinamento Luminoso (ISTIL), Thiene, Italy which supported part of this work. The authors gratefully acknowledge the U.S. Air Force for providing the DMSP data used to make the nighttime lights of the world.
\onecolumn

\begin{figure}
\epsfysize=8cm % fix the y-dimension and scales x-dim. to y-dim.
\hspace{2cm}\epsfbox{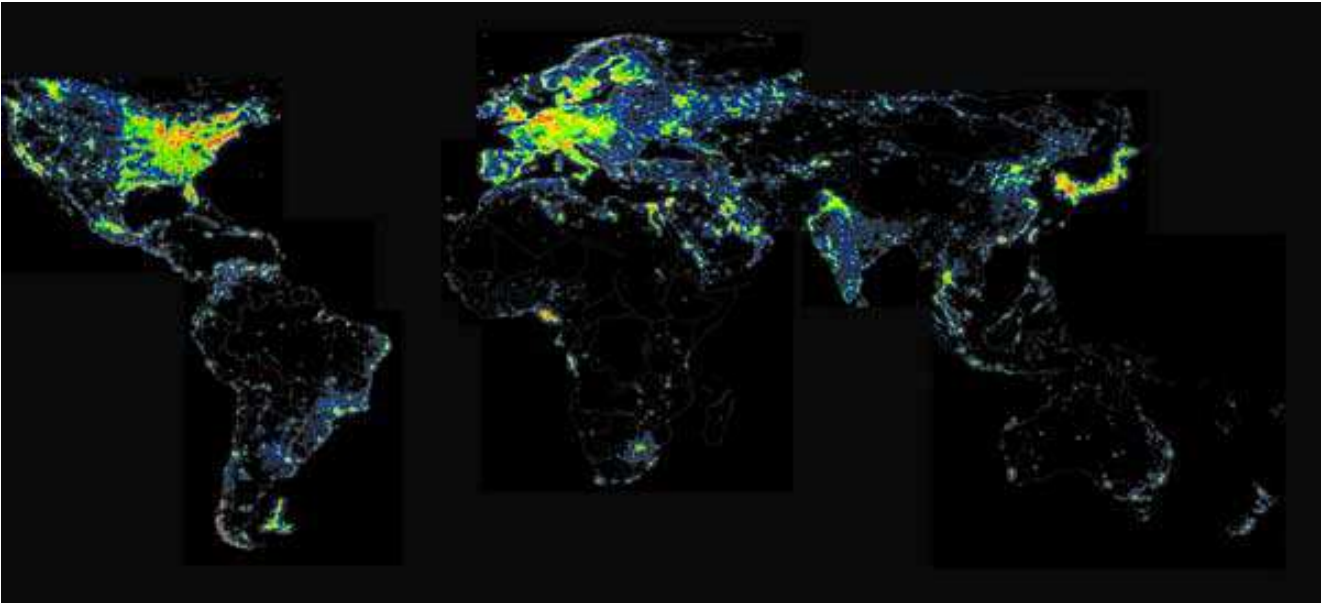} %for centering: act on hspace argument 
\caption[h]{Artificial night sky brightness at sea level in the World. The map has been computed for the photometric astronomical V band, at the zenith, for a clean atmosphere with an aerosol clarity coefficient K=1. The calibration refers to 1996-1997. Country boundaries are approximate.}
\label{f0}
\end{figure}

\begin{figure}
\epsfysize=12cm % fix the y-dimension and scales x-dim. to y-dim.
\hspace{2cm}\epsfbox{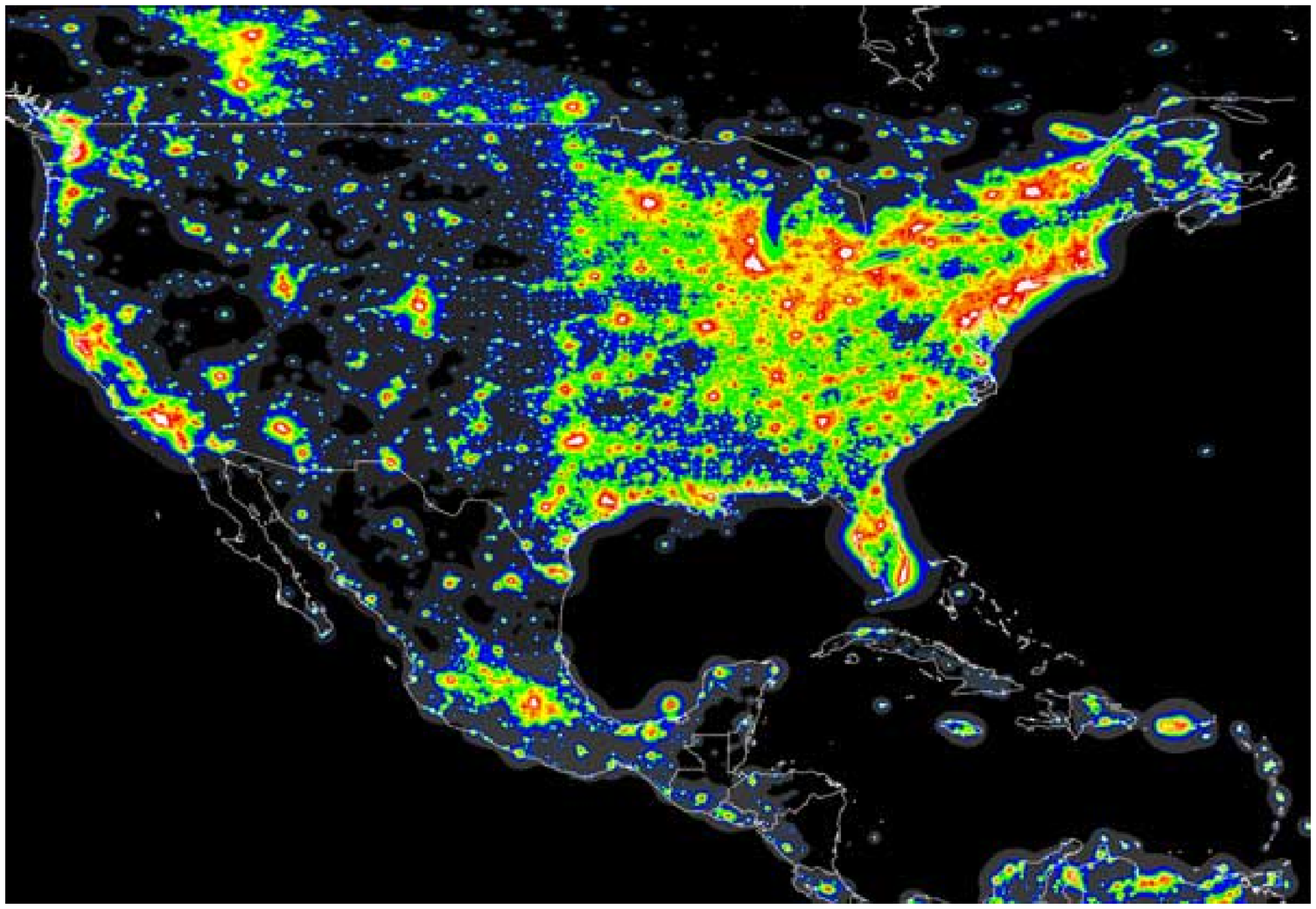} %for centering: act on hspace argument 
\caption[h]{Artificial night sky brightness at sea level for North America. The map has been computed for the photometric astronomical V band, at the zenith, for a clean atmosphere with an aerosol clarity coefficient K=1. The calibration refers to 1996-1997. Country boundaries are approximate.}
\label{f1}
\end{figure}

\begin{figure}
\epsfysize=19cm % fix the y-dimension and scales x-dim. to y-dim.
\hspace{2cm}\epsfbox{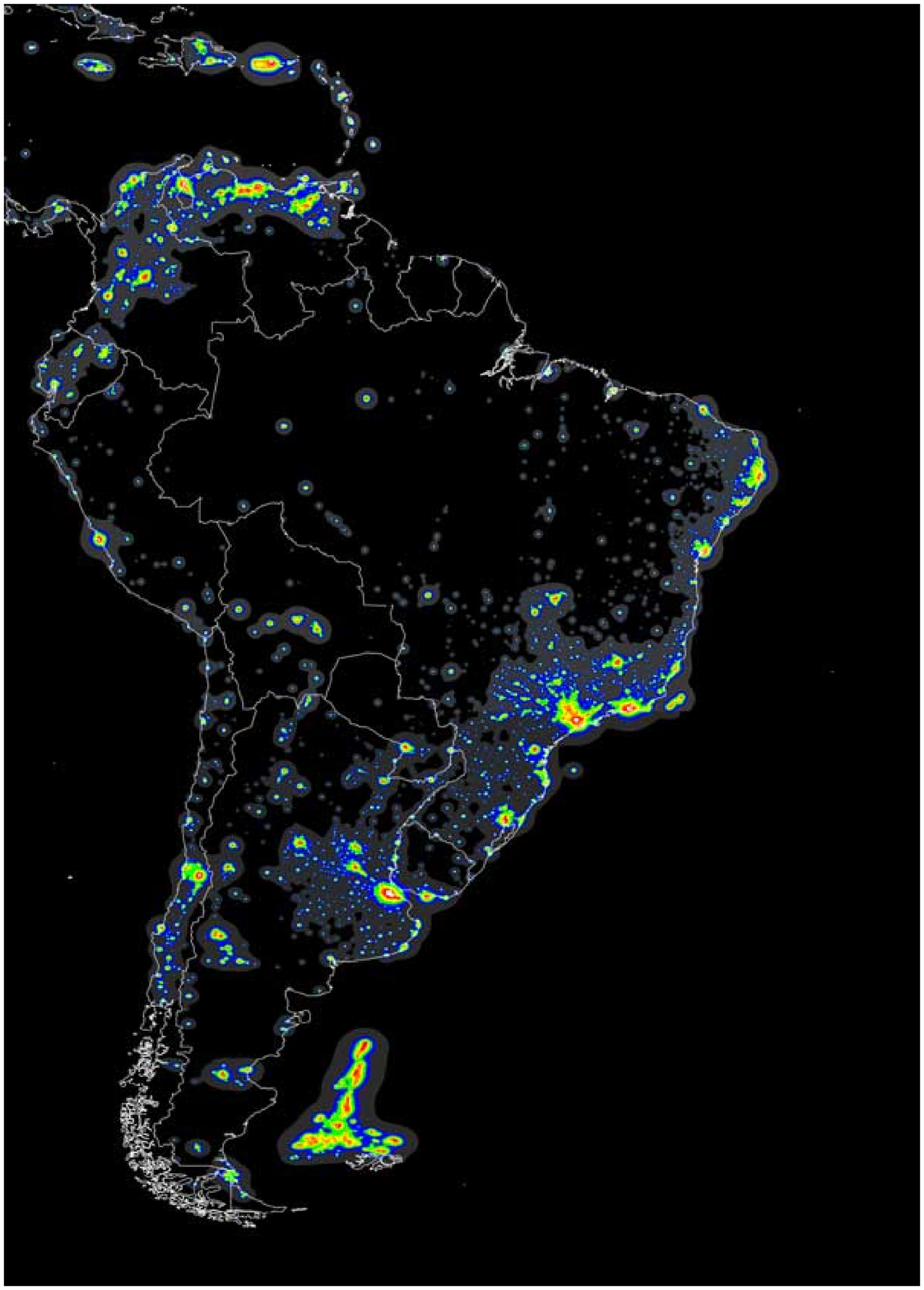} %for centering: act on hspace argument 
\caption[h]{Artificial night sky brightness at sea level for South America. The map has been computed for the photometric astronomical V band, at the zenith, for a clean atmosphere with an aerosol clarity coefficient K=1. The calibration refers to 1996-1997. Country boundaries are approximate.}
\label{f2}
\end{figure}

\begin{figure}
\epsfysize=18cm % fix the y-dimension and scales x-dim. to y-dim.
\hspace{2cm}\epsfbox{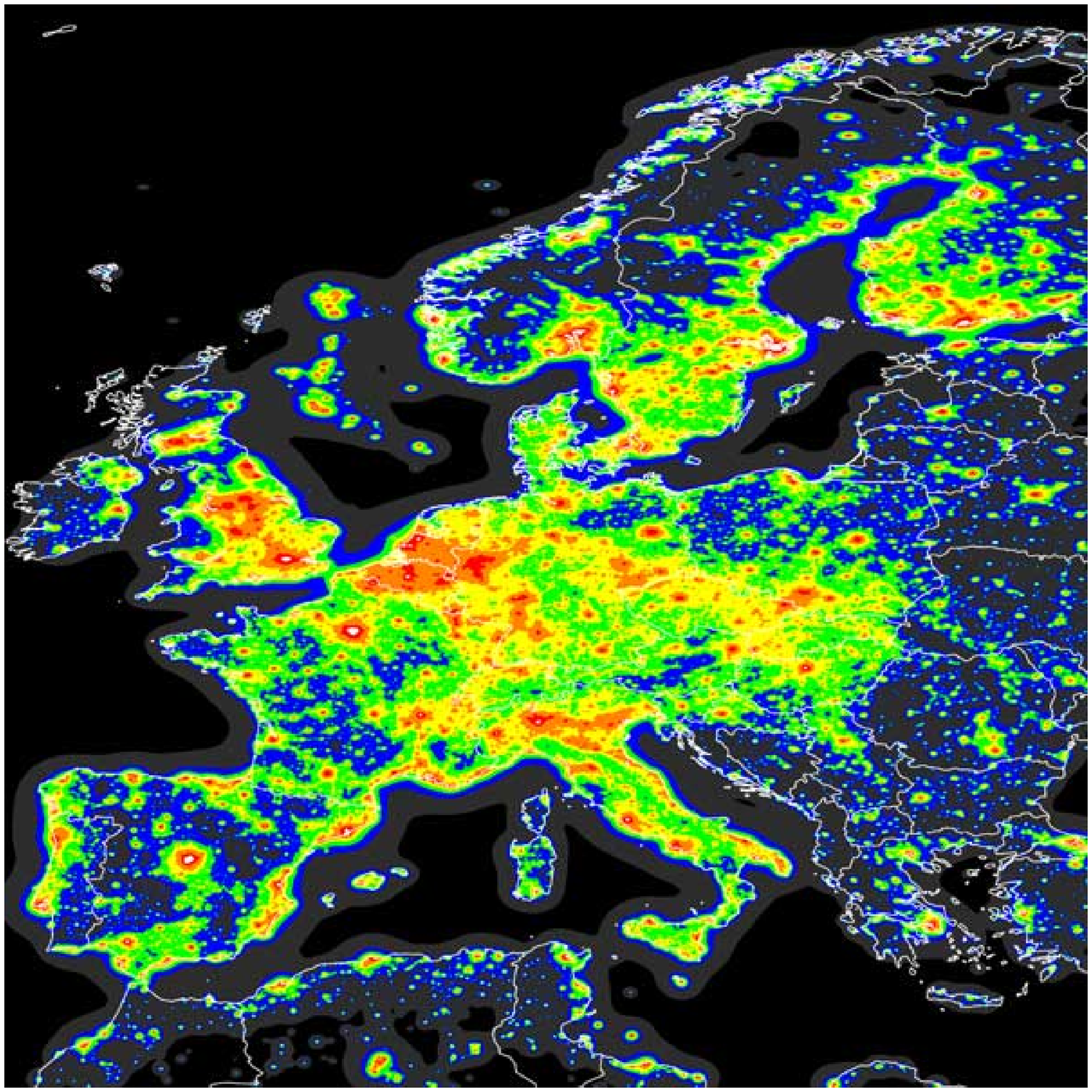} %for centering: act on hspace argument 
\caption[h]{Artificial night sky brightness at sea level for Europe. The map has been computed for the photometric astronomical V band, at the zenith, for a clean atmosphere with an aerosol clarity coefficient K=1. The calibration refers to 1996-1997. Country boundaries are approximate.}
\label{f3}
\end{figure}

\begin{figure}
\epsfysize=18cm % fix the y-dimension and scales x-dim. to y-dim.
\hspace{2cm}\epsfbox{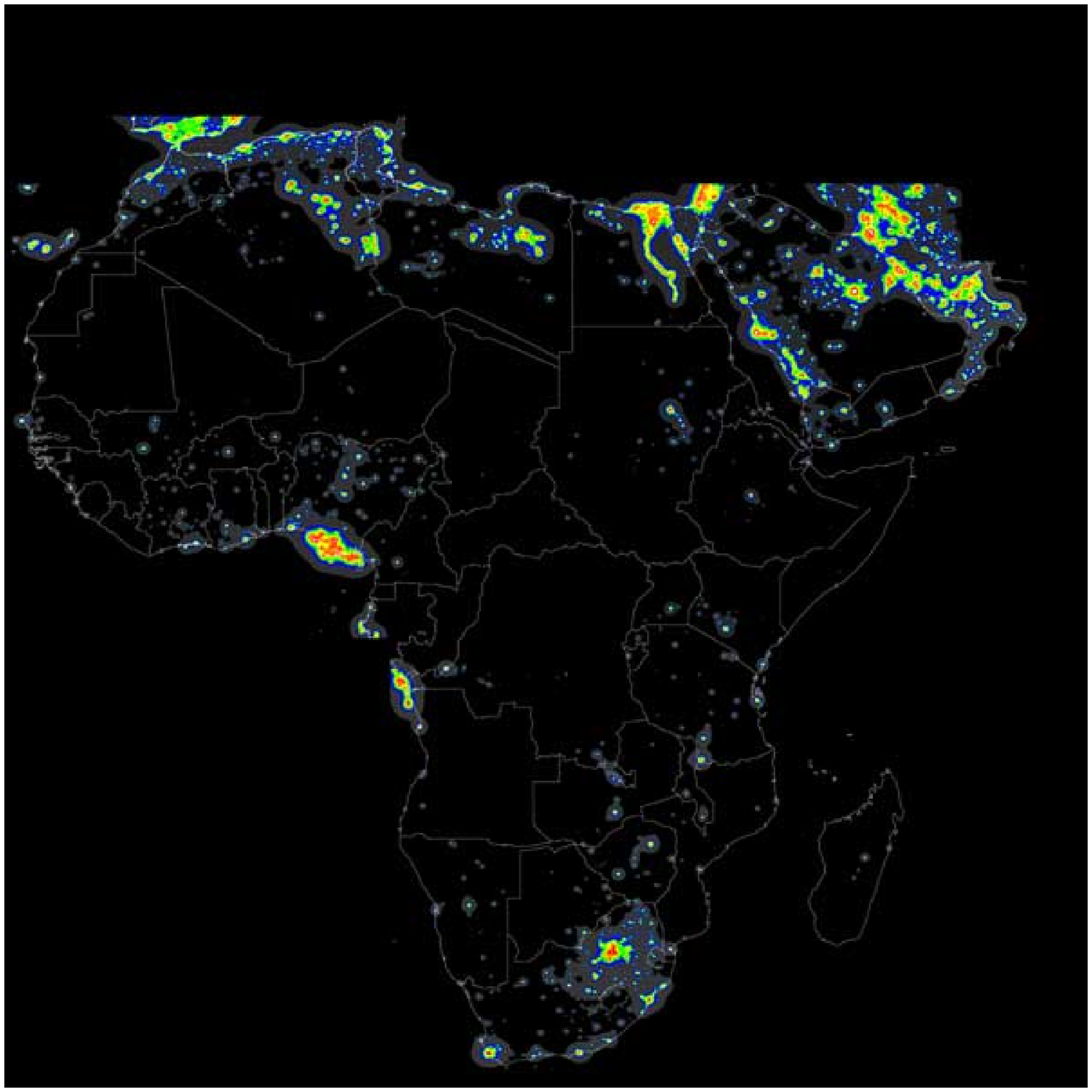} %for centering: act on hspace argument 
\caption[h]{Artificial night sky brightness at sea level for Africa. The map has been computed for the photometric astronomical V band, at the zenith, for a clean atmosphere with an aerosol clarity coefficient K=1. The calibration refers to 1996-1997. Country boundaries are approximate.}
\label{f4}
\end{figure}

\begin{figure}
\epsfysize=19cm % fix the y-dimension and scales x-dim. to y-dim.
\hspace{2cm}\epsfbox{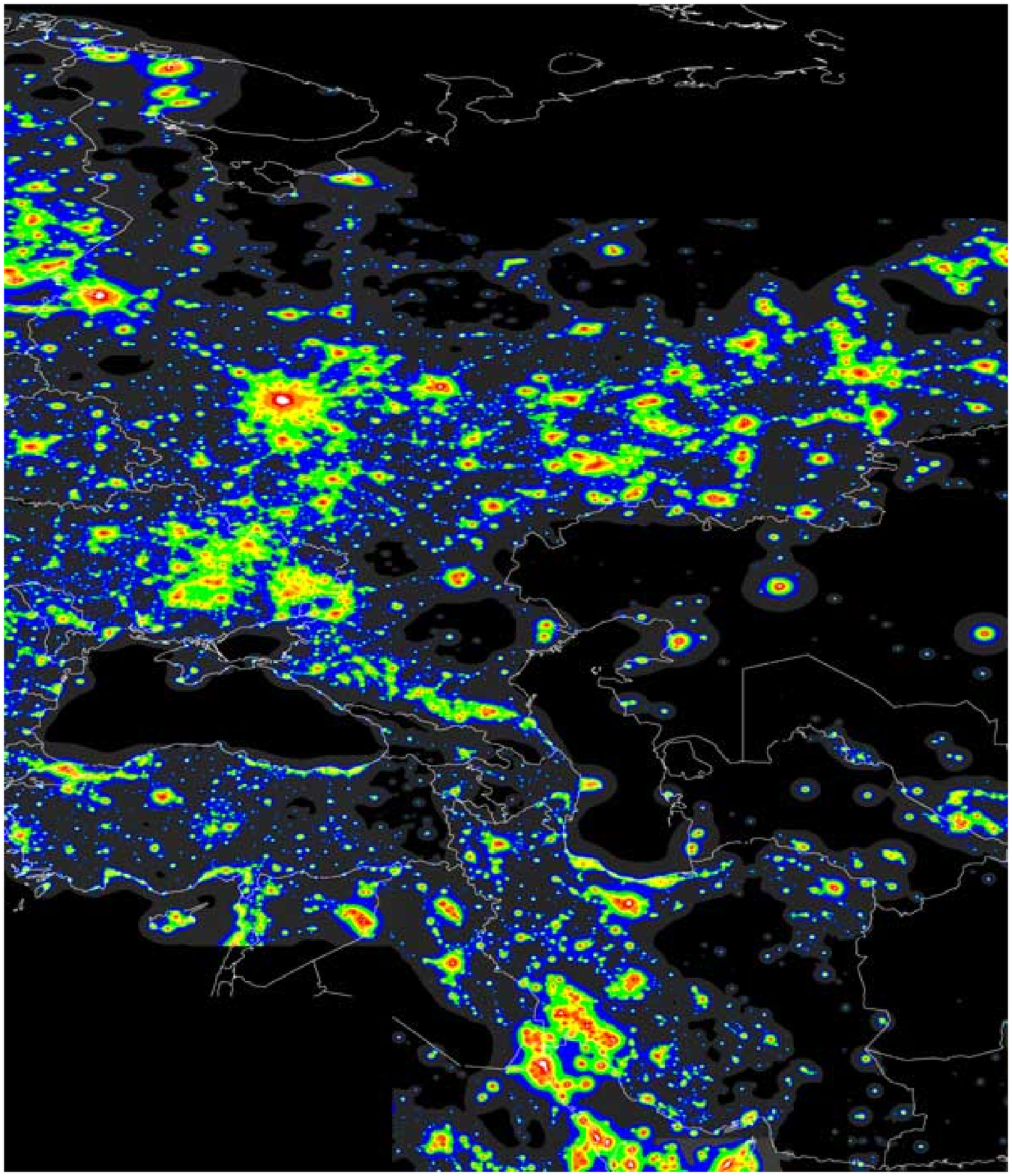} %for centering: act on hspace argument 
\caption[h]{Artificial night sky brightness at sea level for West Asia. The map has been computed for the photometric astronomical V band, at the zenith, for a clean atmosphere with an aerosol clarity coefficient K=1. The calibration refers to 1996-1997. Country boundaries are approximate.}
\label{f5}
\end{figure}

\begin{figure}
\epsfysize=19cm % fix the y-dimension and scales x-dim. to y-dim.
\hspace{2cm}\epsfbox{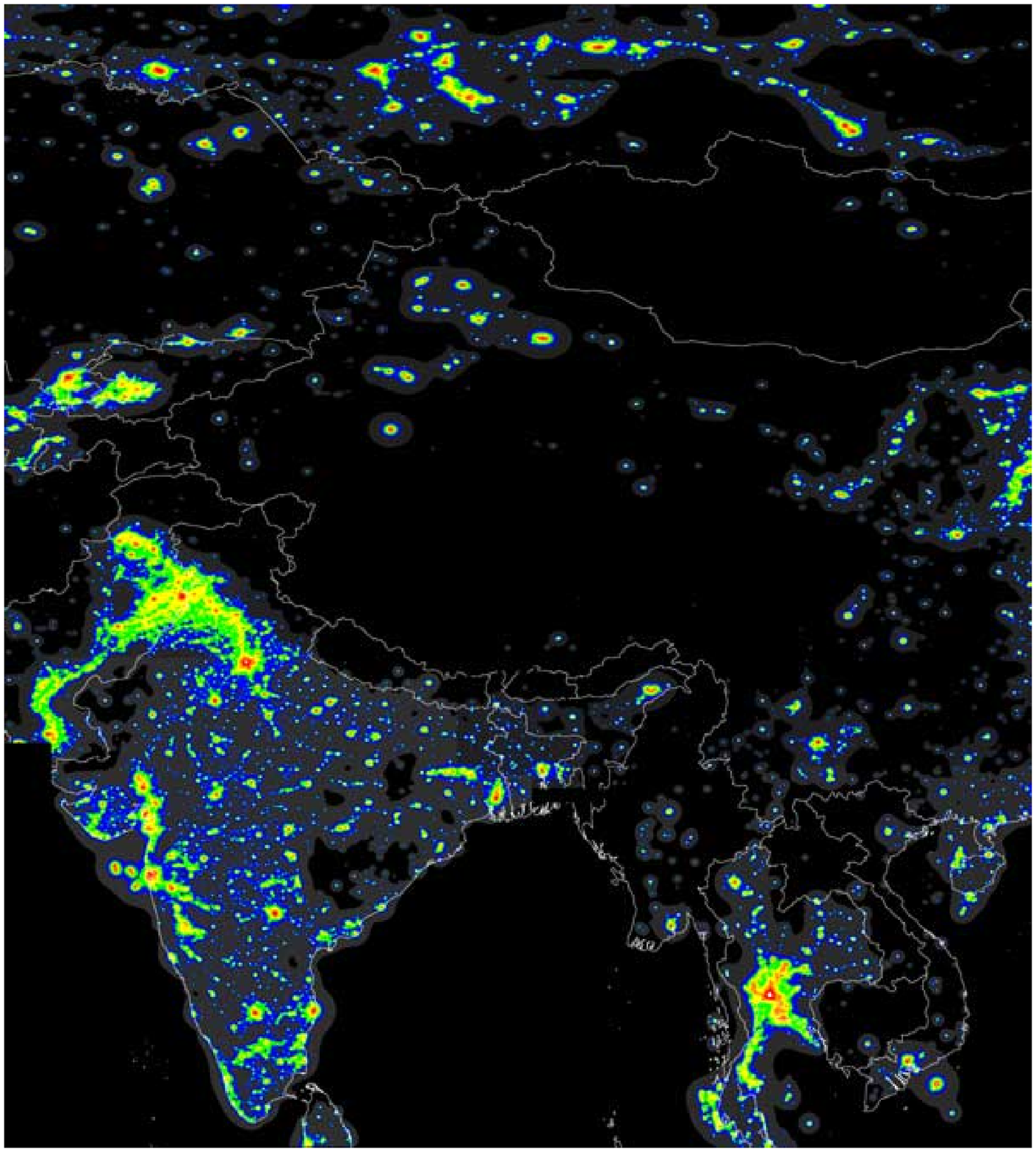} %for centering: act on hspace argument 
\caption[h]{Artificial night sky brightness at sea level for Center Asia. The map has been computed for the photometric astronomical V band, at the zenith, for a clean atmosphere with an aerosol clarity coefficient K=1. The calibration refers to 1996-1997. Country boundaries are approximate.}
\label{f6}
\end{figure}

\begin{figure}
\epsfysize=16cm % fix the y-dimension and scales x-dim. to y-dim.
\hspace{2cm}\epsfbox{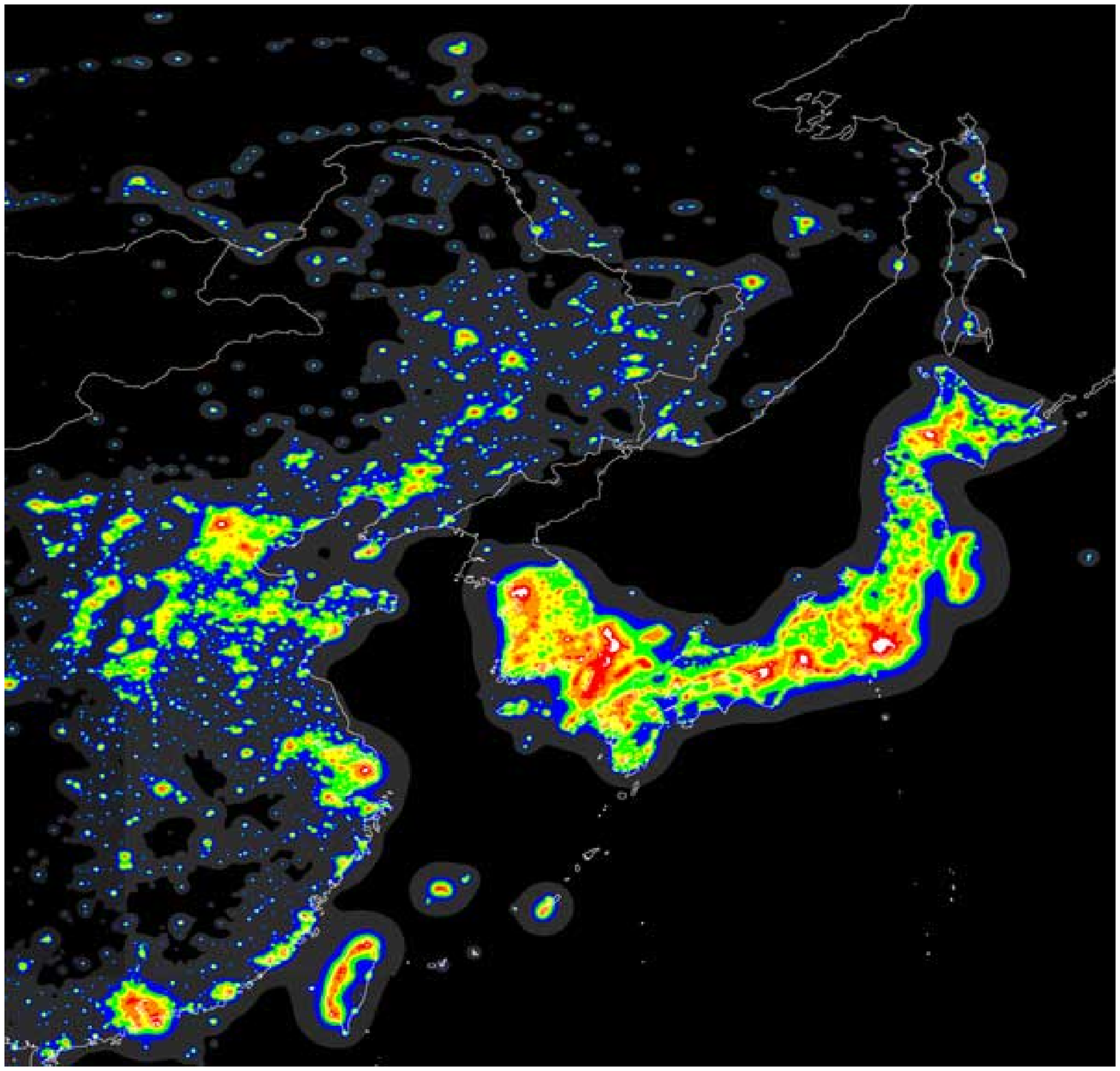} %for centering: act on hspace argument 
\caption[h]{Artificial night sky brightness at sea level for East Asia. The map has been computed for the photometric astronomical V band, at the zenith, for a clean atmosphere with an aerosol clarity coefficient K=1. The calibration refers to 1996-1997. Country boundaries are approximate.}
\label{f7}
\end{figure}

\begin{figure}
\epsfysize=16cm % fix the y-dimension and scales x-dim. to y-dim.
\hspace{2cm}\epsfbox{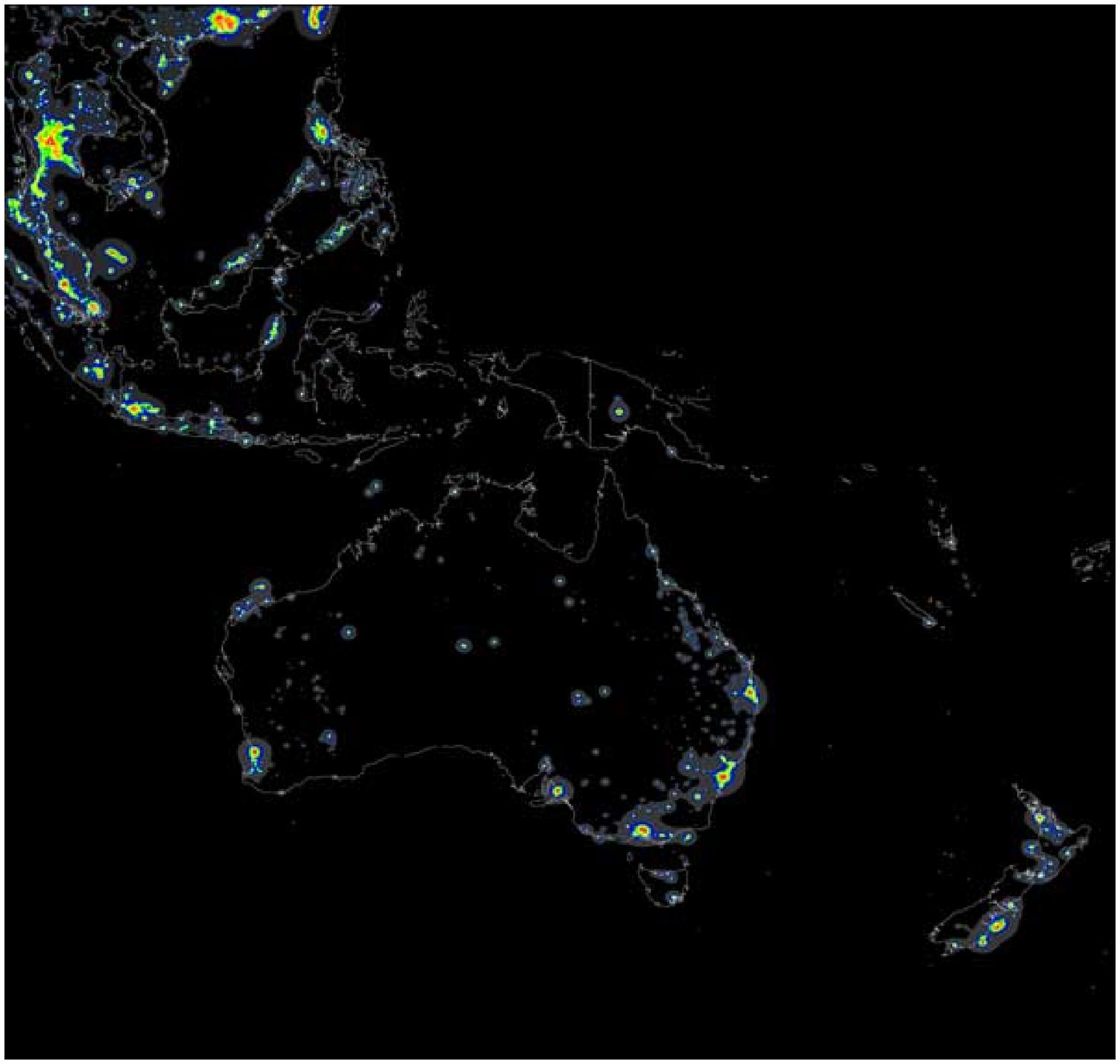} %for centering: act on hspace argument 
\caption[h]{Artificial night sky brightness at sea level for Oceania. The map has been computed for the photometric astronomical V band, at the zenith, for a clean atmosphere with an aerosol clarity coefficient K=1. The calibration refers to 1996-1997. Country boundaries are approximate.}
\label{f8}
\end{figure}

\begin{figure}
\epsfysize=8cm % fix the y-dimension and scales x-dim. to y-dim.
\hspace{-0.1cm}\epsfbox{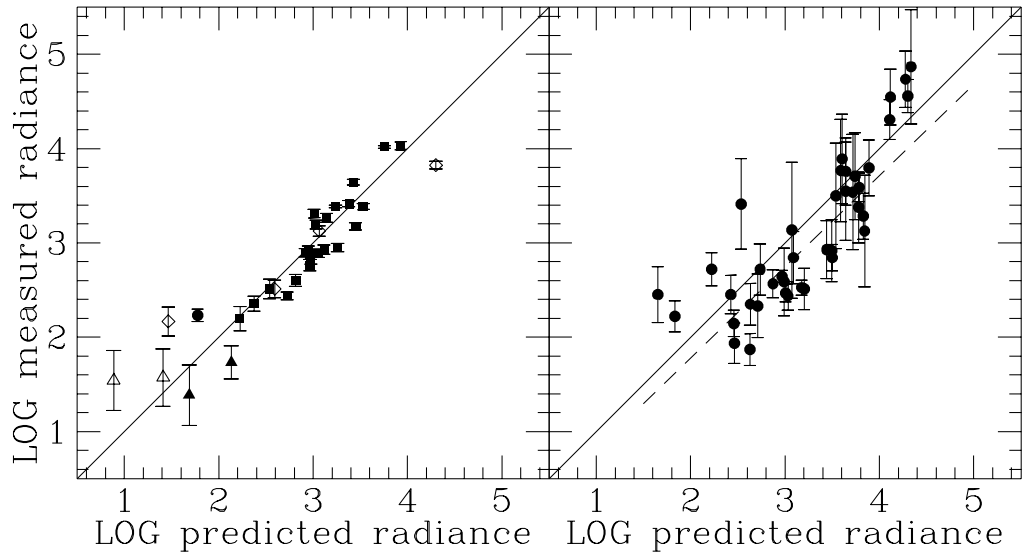} %for centering: act on hspace argument 
\caption[h]{Comparison between map predictions and measurements of artificial night sky brightness.   Left panel: map predictions versus artificial sky brightness measurements at the bottom of the atmosphere taken in clean or photometric nights in the V band in Europe (filled squares), North America (open triangles), South America (open rhombi), Africa (filled triangles), Asia (filled circle).  Right panel: map predictions versus photographic measurements taken in Japan in the period 1987-1991 with variable atmospheric aerosol content and averaged for each site. The large errorbars show the effects of the changes in the atmospheric aerosol content and in the extinction of the light of the comparison star. The dashed line shows the linear regression. Night sky brightnesses are expressed as photon radiances.}
\label{f10}
\end{figure}

\begin{figure}
\epsfysize=9cm % fix the y-dimension and scales x-dim. to y-dim.
\hspace{2cm}\epsfbox{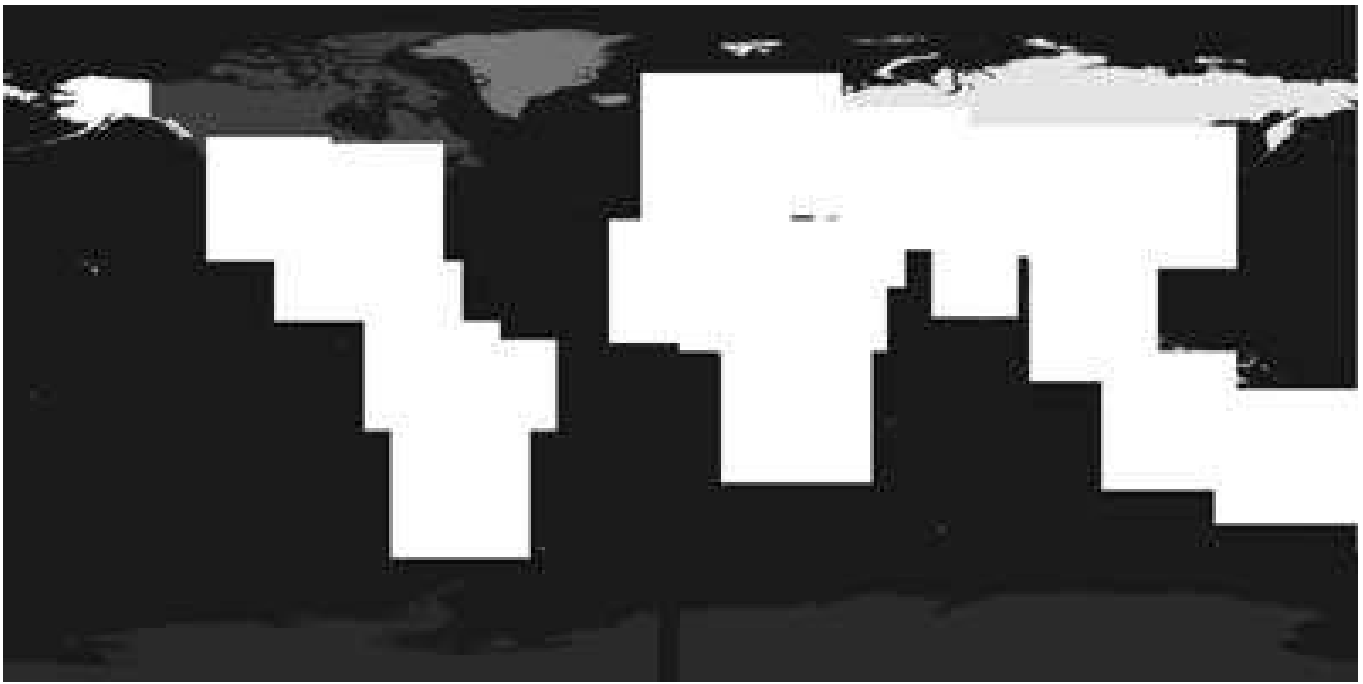} %for centering: act on hspace argument 
\caption[h]{The World areas covered by the Atlas and the statistic (in white).}
\label{f9}
\end{figure}

\twocolumn

{}

\onecolumn

{\tiny

\begin{table*}
\begin{minipage}{200mm}
%  \centering
  \caption{Percentage of population who are living under a sky brightness greater than given levels.}\label{tab1}
  \begin{tabular}{lccccccccccccc}
    \hline
    & (1)&(2)&(3)&(4)&(5)&(6)&(7)&(8)&(9)&(10)&(11)&(12)\\
     Country & $\geq0.11b_{n}$&$\geq0.33b_{n}$&$\geq b_{n}$&$\geq 3b_{n}$&$\geq 9b_{n}$&$\geq 27b_{n}$&$\geq b_{p}$&$\geq b_{fq}$&$\geq b_{m}$&
$\geq b_{fm}$&$\geq b_{mw}$&$\geq b_{e}$\\
     \hline
Afghanistan &   11  &   8   &   1   &   0   &   0   &   0   &   12  &   8   &   1   &   0   &      0   &   0   \\
Albania &   50  &   39  &   27  &   7   &   0   &   0   &   53  &   39  &   27  &   5   &         0   &   0   \\
Algeria &   86  &   74  &   61  &   36  &   12  &   2   &   87  &   73  &   61  &   30  &          16  &   4   \\
Andorra &   100 &   100 &   100 &   90  &   0   &   0   &   100 &   100 &   100 &   85  &          48  &   0   \\
Angola  &   16  &   15  &   14  &   11  &   7   &   0   &   16  &   14  &   14  &   11  &          10  &   0   \\
Anguilla UK &   100 &   99  &   51  &   0   &   0   &   0   &   100 &   99  &   51  &   0   &      0   &   0   \\
Antigua-Barbuda &   98  &   91  &   70  &   21  &   0   &   0   &   98  &   90  &   70  &  0&      0   &   0   \\
Argentina   &   74  &   71  &   67  &   59  &   44  &   23  &   75  &   70  &   67  &   58  &      52  &   29  \\
Armenia &   91  &   88  &   61  &   42  &   0   &   0   &   92  &   88  &   61  &   35  &          0   &   0   \\
Australia   &   71  &   69  &   68  &   62  &   37  &   1   &   71  &   69  &   68  &   60  &      48  &   8   \\
Austria &   100 &   97  &   82  &   45  &   21  &   0   &   100 &   97  &   82  &   41  &          29  &   9   \\
Azerbaigian &   82  &   76  &   54  &   29  &   1   &   0   &   82  &   75  &   54  &   27  &      19  &   0   \\
Bahamas &   84  &   82  &   81  &   75  &   58  &   0   &   85  &   82  &   81  &   73  &          66  &   0   \\
Bahrain &   100 &   100 &   100 &   99  &   99  &   76  &   100 &   100 &   100 &   99  &          99  &   98  \\
Bangladesh  &   43  &   29  &   18  &   8   &   4   &   0   &   45  &   29  &   18  &   8   &     6   &   0   \\
Barbados    &   100 &   98  &   91  &   61  &   0   &   0   &   100 &   98  &   91  &   56  &      27  &   0   \\
Belgium &   100 &   100 &   100 &   96  &   52  &   8   &   100 &   100 &   100 &   94  &          76  &   21  \\
Belize  &   31  &   17  &   6   &   0   &   0   &   0   &   34  &   16  &   6   &   0   &           0   &   0   \\
Benin   &   28  &   24  &   17  &   2   &   0   &   0   &   29  &   24  &   17  &   0   &          0   &   0   \\
Bermuda UK  &   100 &   100 &   100 &   44  &   0   &   0   &   100 &   100 &   100 &   14  &       0   &   0   \\
Bhutan  &   10  &   7   &   0   &   0   &   0   &   0   &   10  &   6   &   0   &   0   &          0   &   0   \\
Bolivia &   58  &   57  &   56  &   52  &   37  &   0   &   58  &   57  &   56  &   50  &           45  &   15  \\
Bosnia-Herzegovina  &   80  &   61  &   32  &   0   &   0   &   0   &   82  &   60  &   32&0&       0   &   0   \\
Botswana    &   23  &   20  &   16  &   8   &   0   &   0   &   23  &   20  &   16  &   8   &       1   &   0   \\
Brazil  &   66  &   60  &   55  &   45  &   29  &   10  &   66  &   59  &   55  &   43  &           36  &   17  \\
British Virgin Islands  &   70  &   63  &   52  &   0   &   0   &   0   & 70& 63& 52  &   0   &      0   &   0   \\
Brunei  &   93  &   84  &   78  &   58  &   28  &   0   &   94  &   84  &   78  &   53  &           44  &   0   \\
Bulgaria    &   84  &   71  &   53  &   22  &   1   &   0   &   86  &   70  &   53  &   19  &       8   &   0   \\
Burkina Faso    &   7   &   6   &   5   &   3   &   0   &   0   &   7   &   6   &   5   &   3 &     0   &   0   \\
Burundi &   6   &   4   &   4   &   0   &   0   &   0   &   6   &   4   &   4   &   0   &            0   &   0   \\
Byelarus    &   86  &   78  &   68  &   45  &   13  &   0   &   87  &   78  &   68  &   37  &       18  &   0   \\
Caiman Islands UK   &   92  &   90  &   78  &   55  &   0   &   0   &   92  &   86 & 78& 50  &     16  &   0   \\
Cambodia    &   14  &   11  &   9   &   7   &   0   &   0   &   15  &   11  &   9   &   5   &       0   &   0   \\
Cameroon    &   22  &   20  &   18  &   14  &   0   &   0   &   22  &   20  &   18  &   9   &       0   &   0   \\
Canada  &   97  &   94  &   90  &   83  &   71  &   46  &   97  &   94  &   90  &   82  &           77  &   59  \\
Central African Rep.    &   0   &   0   &   0   &   0   &   0   & 0  & 0 & 0 & 0   &   0   &       0   &   0   \\
Chad    &   7   &   5   &   5   &   0   &   0   &   0   &   7   &   5   &   5   &   0   &          0   &   0   \\
Chile   &   87  &   83  &   79  &   72  &   49  &   26  &   88  &   82  &   79  &   71  &          61  &   34  \\
China   &   54  &   41  &   29  &   13  &   5   &   1   &   55  &   40  &   29  &   12  &          7   &   2   \\
Cisgiordania    &   100 &   100 &   100 &   81  &   37  &   0   &   100 &  100 & 100 &  73 &       51  &   24  \\
Colombia    &   77  &   68  &   60  &   49  &   34  &   8   &   78  &   67  &   60  &   47  &      40  &   22  \\
Congo       &   41  &   39  &   36  &   33  &   0   &   0   &   41  &   39  &   36  &   27  &      0   &   0   \\
Costa Rica  &   80  &   70  &   64  &   56  &   39  &   0   &   81  &   70  &   64  &   55  &      50  &   14  \\
Croatia &   96  &   85  &   67  &   31  &   15  &   0   &   96  &   84  &   67  &   25  &          17  &   0   \\
Cuba    &   55  &   47  &   39  &   19  &   2   &   0   &   57  &   47  &   39  &   17  &          11  &   0   \\
Cyprus  &   98  &   91  &   83  &   66  &   36  &   0   &   98  &   91  &   83  &   65  &          56  &   0   \\
Czech Republic  &   100 &   100 &   95  &   59  &   22  &   0   &   100 &   100 &  95 & 52  &      34  &   5   \\
Denmark &   100 &   97  &   85  &   53  &   23  &   0   &   100 &   97  &   85  &   50  &          33  &   3   \\
Djibouti    &   24  &   22  &   21  &   14  &   0   &   0   &   25  &   22  &   21  &   0   &      0   &   0   \\
Dominica    &   7   &   4   &   0   &   0   &   0   &   0   &   7   &   4   &   0   &   0   &      0   &   0   \\
Dominican Republic  &   84  &   74  &   64  &   48  &   36  &   0  & 85  & 74  & 64  & 45  &      38  &   16  \\
Ecuador &   57  &   48  &   41  &   27  &   9   &   0   &   58  &   48  &   41  &   24  &        15  &   2   \\
Egypt   &   100 &   100 &   99  &   82  &   33  &   19  &   100 &   100 &   99  &   73  &          43  &   23  \\
El Salvador &   83  &   71  &   55  &   38  &   26  &   0   &   84  &   70  &   55  &   35  &      29  &   5   \\
Equatorial Guinea   &   18  &   15  &   15  &   14  &   5   &   0   &  18  & 15 & 15 & 14  &      13  &   1   \\
Eritrea &   17  &   15  &   13  &   3   &   0   &   0   &   18  &   15  &   13  &   0   &          0   &   0   \\
Estonia &   86  &   72  &   65  &   55  &   35  &   0   &   88  &   72  &   65  &   53  &          42  &   20  \\
Ethiopia    &   6   &   5   &   4   &   4   &   0   &   0   &   6   &   5   &   4   &   4   &      2   &   0   \\
Falkland Islands UK &   8   &   2   &   0   &   0   &   0   &   0   &   9   &  2 & 0 &  0   &      0   &   0   \\
Faroe Islands   &   75  &   69  &   58  &   0   &   0   &   0   &   76  &   69  &   58 & 0  &      0   &   0   \\
Fiji Islands    &   18  &   14  &   1   &   0   &   0   &   0   &   19  &   14  & 1 & 0 &           0   &   0   \\
Finland &   98  &   94  &   88  &   80  &   65  &   24  &   98  &   94  &   88  &   78  &           72  &   44  \\
France  &   100 &   95  &   84  &   67  &   41  &   12  &   100 &   95  &   84  &   64  &          51  &   22  \\
  \hline
  \end{tabular}
  \end{minipage}
\end{table*}

\begin{table*}
\begin{minipage}{200mm}
%  \centering
  \contcaption{}\label{tab2}
  \begin{tabular}{lccccccccccccc}
    \hline
     &(1)&(2)&(3)&(4)&(5)&(6)&(7)&(8)&(9)&(10)&(11)&(12)\\
      Country & $\geq0.11b_{n}$&$\geq0.33b_{n}$&$\geq b_{n}$&$\geq 3b_{n}$&$\geq 9b_{n}$&$\geq 27b_{n}$&$\geq b_{p}$&$\geq b_{fq}$&$\geq b_{m}$&
$\geq b_{fm}$&$\geq b_{mw}$&$\geq b_{e}$\\ \hline
French Guiana   &   37  &   37  &   24  &   0   &   0   &   0   & 37  & 37  & 24  & 0   &      0   &   0   \\
Gabon   &   39  &   37  &   34  &   31  &   1   &   0   &   39  &   37  &   34  &   22  &      1   &   0   \\
Gambia  &   28  &   26  &   23  &   0   &   0   &   0   &   28  &   26  &   23  &   0   &      0   &   0   \\
Gaza    &   100 &   100 &   100 &   95  &   0   &   0   &   100 &   100 &   100 &   79  &      36  &   0   \\
Georgia &   81  &   76  &   50  &   14  &   0   &   0   &   81  &   76  &   50  &   10  &      0   &   0   \\
Germany &   100 &   100 &   94  &   66  &   25  &   0   &   100 &   100 &   94  &   60  &      40  &   5   \\
Ghana   &   29  &   23  &   18  &   12  &   4   &   0   &   30  &   23  &   18  &   11  &      7   &   0   \\
Gibraltar UK    &   100 &   100 &   100 &   100 &   0   &   0   & 100 & 100 & 100 & 100 &      84  &   0   \\
Greece  &   90  &   80  &   70  &   54  &   41  &   17  &   91  &   80  &   70  &   52  &      44  &   31  \\
Grenada &   47  &   43  &   14  &   0   &   0   &   0   &   47  &   42  &   14  &   0   &      0   &   0   \\
Guadeloupe  &   95  &   95  &   88  &   38  &   1   &   0   &   95  & 95  & 88  &   32  &      17  &   0   \\
Guatemala   &   53  &   39  &   30  &   22  &   17  &   0   &   55  & 38  & 30  &   22  &      20  &   3   \\
Guernsey UK &   100 &   100 &   99  &   14  &   0   &   0   &   100 & 100 & 99  &   0   &      0   &   0   \\
Guinea  &   10  &   9   &   7   &   0   &   0   &   0   &   10  &   9   &   7   &   0   &      0   &   0   \\
Guinea-Bissau   &   21  &   18  &   5   &   0   &   0   &   0   &   21  & 18 & 5  & 0   &      0   &   0   \\
Guyana  &   39  &   36  &   32  &   8   &   0   &   0   &   39  &   36  &   32  &   0   &      0   &   0   \\
Haiti   &   24  &   22  &   20  &   16  &   0   &   0   &   25  &   22  &   20  &   14  &      0   &   0   \\
Honduras    &   49  &   41  &   35  &   27  &   12  &   0   &   50  &   41   & 35 & 27  &      22  &   0   \\
Hungary &   100 &   95  &   76  &   41  &   19  &   5   &   100 &   94  &   76  &   37  &      23  &   12  \\
India   &   61  &   41  &   25  &   12  &   4   &   0   &   63  &   40  &   25  &   10  &      6   &   1   \\
Indonesia   &   42  &   33  &   24  &   12  &   4   &   0   &   42  &   33  & 24  & 11  &      6   &   0   \\
Iran    &   88  &   81  &   73  &   57  &   35  &   14  &   89  &   81  &   73  &   54  &      42  &   21  \\
Iraq    &   86  &   77  &   68  &   44  &   24  &   5   &   87  &   77  &   68  &   40  &      28  &   16  \\
Ireland &   86  &   65  &   52  &   37  &   19  &   0   &   88  &   65  &   52  &   34  &      27  &   0   \\
Isle of Man UK  &   100 &   86  &   54  &   0   &   0   &   0   & 100 & 85  & 54  & 0   &      0   &   0   \\
Israel  &   100 &   100 &   99  &   97  &   79  &   26  &   100 &   100 &   99  &   95  &      90  &   52  \\
Italy   &   100 &   99  &   95  &   78  &   35  &   6   &   100 &   99  &   95  &   72  &      50  &   15  \\
Ivory Coast &   26  &   22  &   18  &   14  &   1   &   0   &   26  &   21 & 18  &  13  &      10  &   0   \\
Jamaica &   98  &   87  &   67  &   44  &   26  &   0   &   99  &   85  &   67  &   43  &      33  &   5   \\
Japan   &   100 &   99  &   96  &   86  &   63  &   27  &   100 &   99  &   96  &   84  &      73  &   41  \\
Jersey UK   &   100 &   100 &   96  &   0   &   0   &   0   &   100 &   100 & 96  & 0   &      0   &   0   \\
Jordan  &   94  &   91  &   88  &   70  &   35  &   8   &   94  &   91  &   88  &   65  &      57  &   23  \\
Kazakhstan  &   58  &   54  &   47  &   31  &   3   &   0   &   58  &   54  & 47  & 26  &      9   &   0   \\
Kenya   &   19  &   16  &   12  &   7   &   0   &   0   &   19  &   16  &   12  &   6   &      1   &   0   \\
Kuwait  &   100 &   100 &   100 &   99  &   98  &   86  &   100 &   100 &   100 &   99  &      98  &   96  \\
Kyrgyzstan  &   75  &   66  &   47  &   17  &   0   &   0   &   75  &   66  & 47  & 12  &      6   &   0   \\
Laos    &   16  &   14  &   11  &   2   &   0   &   0   &   16  &   14  &   11  &   0   &     0   &   0   \\
Latvia  &   77  &   68  &   61  &   47  &   33  &   0   &   78  &   68  &   61  &   44  &      37  &   11  \\
Lebanon &   100 &   97  &   81  &   43  &   2   &   0   &   100 &   96  &   81  &   37  &      14  &   0   \\
Lesotho &   17  &   11  &   9   &   0   &   0   &   0   &   17  &   11  &   9   &   0   &      0   &   0   \\
Liberia &   0   &   0   &   0   &   0   &   0   &   0   &   0   &   0   &   0   &   0   &      0   &   0   \\
Libya   &   83  &   78  &   73  &   62  &   39  &   8   &   84  &   78  &   73  &   58  &      48  &   21  \\
Liechtenstein   &   100 &   100 &   100 &   0   &   0   &   0   & 100 & 100 & 100 & 0   &      0   &   0   \\
Lithuania   &   86  &   71  &   61  &   42  &   12  &   0   &   89  &   71  & 61  & 38  &      23  &   0   \\
Luxembourg  &   100 &   100 &   100 &   94  &   63  &   0   &   100 &   100 & 100 & 90  &      81  &   11  \\
Macau P &   100 &   100 &   100 &   100 &   100 &   0   &   100 &   100 &   100 &   100 &      100 &   0   \\
Macedonia   &   92  &   82  &   71  &   32  &   0   &   0   &   94  &   81  & 71  & 29  &      12  &   0   \\
Madagascar  &   11  &   10  &   8   &   0   &   0   &   0   &   11  &   10  & 8   & 0   &      0   &   0   \\
Malawi  &   15  &   13  &   10  &   0   &   0   &   0   &   15  &   13  &   10  &   0   &      0   &   0   \\
Malaysia    &   78  &   68  &   58  &   40  &   20  &   1   &   78  &   68  & 58  & 36  &      25  &   8   \\
Mali    &   15  &   13  &   9   &   6   &   0   &   0   &   15  &   13  &   9   &   5   &      0   &   0   \\
Malta   &   100 &   100 &   100 &   91  &   48  &   0   &   100 &   100 &   100 &   88  &      77  &   0   \\
Martinique F    &   100 &   99  &   93  &   61  &   0   &   0   & 100 & 99  & 93  & 58  &     39  &   0   \\
Mauritania  &   23  &   22  &   21  &   17  &   0   &   0   &   23  & 22  & 21  &   17  &      0   &   0   \\
Mayotte F   &   6   &   5   &   0   &   0   &   0   &   0   &   7   & 3   & 0   &   0   &      0   &   0   \\
Mexico  &   88  &   81  &   72  &   59  &   44  &   25  &   89  &   80  &   72  &   57  &     50  &   33  \\
Moldova &   93  &   83  &   62  &   33  &   0   &   0   &   95  &   83  &   62  &   27  &      9   &   0   \\
Monaco  &   100 &   100 &   100 &   100 &   87  &   0   &   100 &   100 &   100 &   100 &      100 &   0   \\
Mongolia    &   34  &   33  &   31  &   16  &   0   &   0   &   34  & 33  &   31  & 11  &      0   &   0   \\
Montenegro  &   83  &   74  &   58  &   9   &   0   &   0   &   84  & 74  &   58  & 0   &      0   &   0   \\
Montserrat UK   &   56  &   25  &   0   &   0   &   0   &   0   &   59  & 25  & 0 & 0   &      0   &   0   \\
Morocco &   62  &   53  &   45  &   35  &   16  &   4   &   63  &   52  &   45  &   32  &      25  &   7   \\

 \hline
  \end{tabular}
  \end{minipage}
\end{table*}

\begin{table*}
\begin{minipage}{200mm}
%  \centering
  \contcaption{}\label{tab3}
  \begin{tabular}{lccccccccccccc}
    \hline
     &(1)&(2)&(3)&(4)&(5)&(6)&(7)&(8)&(9)&(10)&(11)&(12)\\
      Country & $\geq0.11b_{n}$&$\geq0.33b_{n}$&$\geq b_{n}$&$\geq 3b_{n}$&$\geq 9b_{n}$&$\geq 27b_{n}$&$\geq b_{p}$&$\geq b_{fq}$&$\geq b_{m}$&
$\geq b_{fm}$&$\geq b_{mw}$&$\geq b_{e}$\\ \hline
Mozambique  &   10  &   9   &   7   &   4   &   0   &   0   &   10  &   9   &   7   &   3   &                   1   &   0   \\
Myanmar &   25  &   19  &   12  &   8   &   2   &   0   &   25  &   19  &   12  &   7   &                       5   &   0   \\
Namibia &   17  &   16  &   13  &   8   &   3   &   0   &   17  &   16  &   13  &   8   &                       6   &   0   \\
Nepal   &   25  &   19  &   9   &   3   &   0   &   0   &   26  &   18  &   9   &   2   &                       0   &   0   \\
Netherlands &   100 &   100 &   100 &   88  &   39  &   2   &   100 &   100 &   100 &   85  &                  60  &   16  \\
Netherlands Antilles    &   100 &   98  &   93  &   89  &   56  &   0   &   100 &   98  &   93  &   84  &      69  &   0   \\
New Caledonia   &   45  &   44  &   44  &   42  &   0   &   0   &   45  &   44  &   44  &   41  &               0   &   0   \\
New Zealand &   87  &   84  &   81  &   67  &   25  &   0   &   87  &   84  &   81  &   61  &                  45  &   2   \\
Nicaragua   &   56  &   48  &   42  &   22  &   11  &   0   &   57  &   48  &   42  &   22  &                 20  &   0   \\
Niger   &   3   &   2   &   1   &   1   &   0   &   0   &   3   &   2   &   1   &   1   &                      0   &   0   \\
Nigeria &   45  &   37  &   27  &   17  &   7   &   1   &   46  &   36  &   27  &   15  &                     12  &   2   \\
Norfolk Islands AU  &   7   &   0   &   0   &   0   &   0   &   0   &   10  &   0   &   0   &   0   &           0   &   0   \\
North Korea &   25  &   18  &   13  &   1   &   0   &   0   &   26  &   17  &   13  &   0   &                   0   &   0   \\
Norway  &   95  &   89  &   82  &   72  &   52  &   20  &   96  &   89  &   82  &   70  &                      61  &   30  \\
Oman    &   90  &   83  &   73  &   39  &   24  &   0   &   90  &   82  &   73  &   35  &                      27  &   12  \\
Pakistan    &   87  &   77  &   54  &   26  &   14  &   4   &   88  &   76  &   54  &   24  &                 18  &   8   \\
Panama  &   65  &   57  &   49  &   38  &   23  &   0   &   65  &   57  &   49  &   36  &                      29  &   0   \\
Papua New Guinea    &   13  &   12  &   10  &   3   &   0   &   0   &   13  &   12  &   10  &   2   &           0   &   0   \\
Paraguay    &   60  &   55  &   50  &   41  &   31  &   0   &   61  &   55  &   50  &   38  &                  36  &   16  \\
Peru    &   58  &   56  &   52  &   44  &   30  &   15  &   59  &   56  &   52  &   41  &                    33  &   23  \\
Philippines &   50  &   42  &   34  &   23  &   14  &   2   &   50  &   42  &   34  &   22  &                 17  &   8   \\
Poland  &   99  &   88  &   72  &   44  &   18  &   0   &   100 &   87  &   72  &   39  &                      26  &   3   \\
Portugal    &   98  &   90  &   80  &   60  &   33  &   14  &   99  &   89  &   80  &   57  &                  42  &   22  \\
Puerto Rico &   100 &   100 &   100 &   93  &   46  &   23  &   100 &   100 &   100 &   90  &                 67  &   33  \\
Qatar   &   100 &   100 &   99  &   97  &   92  &   81  &   100 &   100 &   99  &   96  &                      94  &   84  \\
Romania &   84  &   69  &   52  &   23  &   7   &   0   &   86  &   69  &   52  &   19  &                      13  &   0   \\
Russia  &   87  &   80  &   73  &   60  &   34  &   8   &   88  &   79  &   73  &   57  &                     44  &   15  \\
Rwanda  &   6   &   4   &   4   &   0   &   0   &   0   &   6   &   4   &   4   &   0   &                      0   &   0   \\
Saint Kitts e Nevis &   97  &   84  &   65  &   0   &   0   &   0   &   99  &   81  &   65  &   0   &          0   &   0   \\
Saint Lucia &   88  &   84  &   69  &   0   &   0   &   0   &   89  &   84  &   69  &   0   &                  0   &   0   \\
San Marino  &   100 &   100 &   100 &   99  &   0   &   0   &   100 &   100 &   100 &   90  &                  0   &   0   \\
Saudi Arabia    &   94  &   92  &   90  &   84  &   74  &   53  &   94  &   92  &   90  &   83  &              78  &   64  \\
Senegal &   35  &   32  &   26  &   18  &   0   &   0   &   35  &   31  &   26  &   18  &                      2   &   0   \\
Serbia  &   95  &   83  &   63  &   22  &   5   &   0   &   96  &   82  &   63  &   19  &                     12  &   0   \\
Seychelles  &   0   &   0   &   0   &   0   &   0   &   0   &   0   &   0   &   0   &   0   &                  0   &   0   \\
Sierra Leone    &   15  &   15  &   14  &   0   &   0   &   0   &   15  &   15  &   14  &   0   &             0   &   0   \\
Singapore   &   100 &   100 &   100 &   100 &   100 &   60  &   100 &   100 &   100 &   100 &                 100 &   95  \\
Slovakia    &   100 &   100 &   92  &   35  &   7   &   0   &   100 &   100 &   92  &   29  &                14  &   0   \\
Slovenia    &   100 &   98  &   81  &   47  &   19  &   0   &   100 &   97  &   81  &   43  &                 30  &   0   \\
Somalia &   11  &   9   &   0   &   0   &   0   &   0   &   11  &   9   &   0   &   0   &                     0   &   0   \\
South Africa    &   58  &   51  &   46  &   38  &   23  &   1   &   58  &   51  &   46  &   36  &             29  &   10  \\
South Korea &   100 &   100 &   99  &   92  &   75  &   45  &   100 &   100 &   99  &   90  &                  82  &   59  \\
Spain   &   98  &   93  &   87  &   78  &   57  &   25  &   99  &   93  &   87  &   76  &                    67  &   38  \\
Sri Lanka   &   44  &   26  &   12  &   0   &   0   &   0   &   46  &   24  &   12  &   0   &                 0   &   0   \\
St.Vinc. - Grenadines   &   77  &   62  &   21  &   0   &   0   &   0   &   78  &   62  &   21  &   0   &      0   &   0   \\
Sudan   &   23  &   21  &   18  &   13  &   8   &   0   &   23  &   20  &   18  &   13  &                     10  &   0   \\
Suriname    &   66  &   59  &   53  &   30  &   0   &   0   &   66  &   59  &   53  &   18  &                 0   &   0   \\
Swaziland   &   22  &   14  &   10  &   0   &   0   &   0   &   23  &   14  &   10  &   0   &                  0   &   0   \\
Sweden  &   99  &   97  &   93  &   79  &   51  &   18  &   99  &   97  &   93  &   75  &                     62  &   31  \\
Switzerland &   100 &   100 &   97  &   67  &   15  &   0   &   100 &   100 &   97  &   57  &                 28  &   0   \\
Syria   &   89  &   79  &   65  &   42  &   13  &   0   &   89  &   78  &   65  &   39  &                      23  &   0   \\
Taiwan  &   100 &   99  &   99  &   92  &   60  &   16  &   100 &   99  &   99  &   87  &                      72  &   34  \\
Tajikistan  &   73  &   61  &   41  &   8   &   0   &   0   &   74  &   60  &   41  &   3   &                  0   &   0   \\
Tanzania    &   14  &   12  &   11  &   6   &   0   &   0   &   14  &   12  &   11  &   6   &                  5   &   0   \\
Thailand    &   68  &   56  &   45  &   25  &   14  &   8   &   69  &   56  &   45  &   22  &                   17  &   11  \\
Togo    &   19  &   17  &   15  &   2   &   0   &   0   &   19  &   17  &   15  &   0   &                      0   &   0   \\
Trinidad and Tobago &   99  &   96  &   90  &   67  &   2   &   0   &   99  &   96  &   90  &   59  &          29  &   0   \\
Tunisia &   80  &   70  &   60  &   38  &   11  &   0   &   82  &   69  &   60  &   33  &                     22  &   2   \\
Turkey  &   79  &   71  &   62  &   40  &   15  &   0   &   80  &   70  &   62  &   36  &                     25  &   2   \\
Turkmenistan    &   56  &   50  &   38  &   19  &   4   &   0   &   57  &   49  &   38  &   16  &             11  &   0   \\
Turks - Caicos Is. UK &   54  &   52  &   0   &   0   &   0   &   0   &   54  &   52  &   0   &   0   &        0   &   0   \\
Uganda  &   10  &   8   &   5   &   4   &   0   &   0   &   10  &   8   &   5   &   4   &      1   &   0   \\

   \hline
  \end{tabular}
  \end{minipage}
\end{table*}
}

{\tiny
\begin{table*}
\begin{minipage}{200mm}
%  \centering
  \contcaption{}\label{tab4}
  \begin{tabular}{lccccccccccccc}
    \hline
       &(1)&(2)&(3)&(4)&(5)&(6)&(7)&(8)&(9)&(10)&(11)&(12)\\
         Country & $\geq0.11b_{n}$&$\geq0.33b_{n}$&$\geq b_{n}$&$\geq 3b_{n}$&$\geq 9b_{n}$&$\geq 27b_{n}$&$\geq b_{p}$&$\geq b_{fq}$&$\geq b_{m}$&
$\geq b_{fm}$&$\geq b_{mw}$&$\geq b_{e}$\\\hline
Ukraine &   93  &   85  &   70  &   40  &   7   &   0   &   93  &   85  &   70  &   34  &      18  &   0   \\
United Arab Emirates    &   100 &   100 &   99  &   97  &   89  &   67  &   100 &   100 &   99  &   97  &      94  &   78  \\
United Kingdom  &   100 &   98  &   94  &   79  &   40  &   4   &   100 &   98  &   94  &   74  &      55  &   15  \\
Un. States of America    &   99  &   97  &   93  &   83  &   62  &   30  &   99  &   97  &   93  &   81  &      71  &   44  \\
Uruguay &   80  &   75  &   73  &   62  &   50  &   18  &   80  &   75  &   73  &   61  &      54  &   37  \\
Uzbekistan  &   90  &   84  &   68  &   28  &   10  &   0   &   90  &   83  &   68  &   24  &      12  &   1   \\
Vanuatu &   8   &   6   &   4   &   4   &   0   &   0   &   8   &   5   &   4   &   4   &      2   &   0   \\
Venezuela   &   90  &   85  &   80  &   71  &   52  &   23  &   91  &   84  &   80  &   70    &   62  &   31  \\
Vietnam &   31  &   22  &   14  &   9   &   4   &   0   &   32  &   22  &   14  &   8   &      5   &   2   \\
Virgin islands US   &   100 &   100 &   99  &   94  &   0   &   0   &   100 &   100 &   99  &   84  &      39  &   0   \\
Western Sahara  &   11  &   9   &   8   &   2   &   0   &   0   &   12  &   9   &   8   &   2   &      0   &   0   \\
Yemen   &   41  &   34  &   23  &   13  &   3   &   0   &   42  &   33  &   23  &   12  &      7   &   0   \\
Zaire   &   13  &   12  &   11  &   7   &   0   &   0   &   13  &   12  &   11  &   7   &      1   &   0   \\
Zambia  &   38  &   36  &   32  &   12  &   0   &   0   &   38  &   36  &   32  &   11  &      4   &   0   \\
Zimbabwe    &   30  &   28  &   25  &   17  &   0   &   0   &   30  &   28  &   25  &   14  &      1   &   0   \\
European Union  &   99  &   97  &   90  &   72  &   38  &   8   &   99  &   96  &   90  &   68  &      51  &   17  \\
The World       &   62  &   53  &   43  &  30   & 16    &   6   &   63  &   52  &   43  &   28  &      21  &    9      \\
 
    \hline
  \end{tabular}
  \end{minipage}
\end{table*}

\begin{table*}
\begin{minipage}{200mm}
%  \centering
  \caption{Numerical values and references of thresholds in table 1, columns 8-14. The natural sky brightness has been subtracted.}\label{tab11}
  \begin{tabular}{cccccc}
    \hline
$b_{p}$& $b_{fq}$& $b_{m}$& $b_{fm}$& $b_{mw}$& $b_{e}$\\
 \hline
10\% $b_{n}$ & $\sim$90$\mu cd/m^{2}$ & 252$\mu cd/m^{2}$ & $\sim$890$\mu cd/m^{2}$ & 6 $b_{n}$ & 4452$\mu cd/m^{2}$\\
Smith 1979 & e.g. Walker 1987 & based on Krisciunas \& Schaefer 1991 & e.g. Walker 1987&  estimate & Garstang 1986\\
 \hline
  \end{tabular}
  \end{minipage}
\end{table*}

\begin{table*}
\begin{minipage}{80mm}
%  \centering
  \caption{Percentage of the surface area where the artificial sky brightness at sea level in standard clear nights is greater than given levels.}\label{tab5}
  \begin{tabular}{lcccccc}
    \hline
         &(1)&(2)&(3)&(4)&(5)&(6)\\
         Country & $\geq 0.11b_{n}$&$\geq 0.33b_{n}$&$\geq b_{n}$&$\geq 3b_{n}$&$\geq 9b_{n}$&$\geq 27b_{n}$\\
         \hline
Afghanistan &   0,4 &   0,1 &   0   &   0   &   0   &   0   \\
Albania &   17,1    &   5,2 &   1,3 &   0,1 &   0   &   0   \\
Algeria &   9,4 &   4,4 &   1,8 &        0,7 &   0,2 &   0   \\
Andorra &   100 &   100 &   89,8    &   27,9    &   0   &   0   \\
Angola  &   0,9 &   0,4 &   0,2 &   0   &   0   &   0   \\
Anguilla UK &   100 &   83,6    &   19  &   0   &   0   &   0   \\
Antigua-Barbuda &   63,5    &   49,8    &   21,6    &   1,3 &   0   &   0   \\
Argentina   &   11,3    &   4,6 &   1,9 &   0,7 &   0,2 &   0   \\
Armenia &   17,8    &   7,2 &   2,1 &   0,5 &   0   &   0   \\
Australia   &   2,3 &   1   &   0,4 &   0,2 &   0   &   0   \\
Austria &   100 &   76,2    &   29,3    &   3,5 &   0,4 &   0   \\
Azerbaigian &   23,3    &   9,3 &   3,2 &   0,8 &   0   &   0   \\
Bahamas &   7,8 &   4,9 &   3,4 &   1,7 &   0,3 &   0   \\
Bahrain &   100 &   100 &   91,6    &   74,6    &   50,7    &   25,8    \\
Bangladesh  &   24,4    &   9   &   3   &   0,6 &   0,1 &   0   \\
Barbados    &   100 &   93,3    &   64,6    &   20  &   0   &   0   \\
Belgium &   100 &   100 &   99,8    &   74,4    &   11,4    &   0,3 \\
Belize  &   7,6 &   2,5 &   0,6 &   0   &   0   &   0   \\
Benin   &   1,6 &   0,6 &   0,2 &   0   &   0   &   0   \\
Bermuda UK  &   100 &   100 &   92,5    &   17  &   0   &   0   \\
Bhutan  &   0,4 &   0,1 &   0   &   0   &   0   &   0   \\
Bolivia &   3   &   1,4 &   0,6 &   0,2 &   0   &   0   \\
Bosnia-Herzegovina  &   40,5    &   12,6    &   2,2 &   0   &   0   &   0   \\
Botswana    &   0,6 &   0,2 &   0,1 &   0   &   0   &   0   \\
Brazil  &   7,9 &   3,5 &   1,4 &   0,5 &   0,1 &   0   \\
British Virgin Islands  &   51  &   44,5    &   32,8    &   0   &   0   &   0   \\
Brunei  &   47,6    &   27,2    &   15,8    &   8,4 &   1,1 &   0   \\
Bulgaria    &   41,1    &   12  &   3,4 &   0,4 &   0   &   0   \\
Burkina Faso    &   0,9 &   0,4 &   0,1 &   0   &   0   &   0   \\
Burundi &   1,6 &   0,6 &   0,2 &   0   &   0   &   0   \\
Byelarus    &   41  &   14,8    &   4,9 &   0,8 &   0,1 &   0   \\
Caiman Islands UK   &   68,8    &   59,3    &   27,5    &   10,9    &   0   &   0   \\
Cambodia    &   1,3 &   0,5 &   0,2 &   0   &   0   &   0   \\
Cameroon    &   1,4 &   0,5 &   0,1 &   0   &   0   &   0   \\
Canada  &   32,8    &   18,6    &   9,2 &   3,6 &   1   &   0,2 \\
Central African Rep.    &   0   &   0   &   0   &   0   &   0   &   0   \\
Chad    &   0,1 &   0   &   0   &   0   &   0   &   0   \\
Chile   &   12,2    &   5,6 &   2,1 &   0,7 &   0,2 &   0   \\
China   &   12,5    &   6   &   2,4 &   0,5 &   0,1 &   0   \\
Cisgiordania    &   100 &   100 &   92,7    &   43,2    &   4,1 &   0   \\
Colombia    &   14  &   5,9 &   2,3 &   0,7 &   0,1 &   0   \\
Congo       &    1,2 &  0,5  & 0,2  &   0 &   0 &  0 \\
Costa Rica  &   34,1    &   15,1    &   6   &   2,1 &   0,5 &   0   \\
Croatia &   74,8    &   41,4    &   14,3    &   1,7 &   0,2 &   0   \\
Cuba    &   14,6    &   5,8 &   2,1 &   0,5 &   0,1 &   0   \\
Cyprus  &   85,1    &   57,3    &   29,5    &   7,1 &   0,7 &   0   \\
Czech Republic  &   100 &   99,7    &   76  &   11,8    &   0,9 &   0   \\
Denmark &   99,5    &   87,4    &   46  &   9,1 &   0,9 &   0   \\
Djibouti    &   1,6 &   0,6 &   0,2 &   0   &   0   &   0   \\
Dominica    &   8,1 &   2,5 &   0   &   0   &   0   &   0   \\
Dominican Republic  &   45  &   22,6    &   8,4 &   2,4 &   0,6 &   0   \\
Ecuador &   17,4    &   8,2 &   3,3 &   0,7 &   0,1 &   0   \\
Egypt   &   17,1    &   10,7    &   6,4 &   2,5 &   0,3 &   0,1 \\
El Salvador &   59,1    &   33,4    &   12,4    &   3,4 &   0,8 &   0   \\
Equatorial Guinea   &   5,5 &   3   &   1,3 &   0,5 &   0,1 &   0   \\
Eritrea &   0,8 &   0,3 &   0,1 &   0   &   0   &   0   \\
Estonia &   59,7    &   23,8    &   9,3 &   2,4 &   0,5 &   0   \\
Ethiopia    &   0,4 &   0,2 &   0,1 &   0   &   0   &   0   \\
Falkland Islands UK &   5,1 &   0   &   0   &   0   &   0   &   0   \\
Faroe Islands   &   33,8    &   13,5    &   2,3 &   0   &   0   &   0   \\

  \hline
  \end{tabular}
  \end{minipage}
\end{table*}

\begin{table*}
\begin{minipage}{80mm}
%  \centering
  \contcaption{}\label{tab6}
  \begin{tabular}{|lcccccc}
    \hline
        &(1)&(2)&(3)&(4)&(5)&(6)\\
             Country & $\geq 0.11b_{n}$&$\geq 0.33b_{n}$&$\geq b_{n}$&$\geq 3b_{n}$&$\geq 9b_{n}$&$\geq 27b_{n}$\\
     \hline
Fiji Islands    &   3   &   1   &   0,1 &   0   &   0   &   0   \\
Finland &   70  &   47,2    &   22,7    &   6,7 &   1,3 &   0,2 \\
France  &   98,9    &   75,1    &   36  &   9,6 &   1,4 &   0,1 \\
French Guiana   &   0,6 &   0,3 &   0,1 &   0   &   0   &   0   \\
Gabon   &   3,3 &   1,5 &   0,8 &   0,3 &   0   &   0   \\
Gambia  &   2,5 &   1,2 &   0,6 &   0   &   0   &   0   \\
Gaza    &   100 &   100 &   100 &   74,5    &   0   &   0   \\
Georgia &   10,6    &   4,5 &   1,2 &   0,1 &   0   &   0   \\
Germany &   100 &   94,5    &   64,5    &   16,9    &   1,9 &   0   \\
Ghana   &   4,3 &   1,7 &   0,7 &   0,3 &   0   &   0   \\
Gibraltar UK    &   100 &   100 &   100 &   100 &   0   &   0   \\
Greece  &   57,7    &   25,7    &   9,4 &   2,2 &   0,6 &   0,1 \\
Grenada &   24,5    &   13,5    &   3,3 &   0   &   0   &   0   \\
Guadeloupe  &   89,5    &   87,5    &   55,6    &   10,3    &   0   &   0   \\
Guatemala   &   16,1    &   5,8 &   2,1 &   0,7 &   0,2 &   0   \\
Guernsey UK &   100 &   100 &   97,7    &   3,8 &   0   &   0   \\
Guinea  &   0,4 &   0,2 &   0,1 &   0   &   0   &   0   \\
Guinea-Bissau   &   1,8 &   0,7 &   0,2 &   0   &   0   &   0   \\
Guyana  &   0,4 &   0,2 &   0,1 &   0   &   0   &   0   \\
Haiti   &   5,1 &   2,1 &   0,9 &   0,3 &   0   &   0   \\
Honduras    &   10,6    &   4,5 &   1,8 &   0,5 &   0   &   0   \\
Hungary &   100 &   81,9    &   29,9    &   3,4 &   0,6 &   0,1 \\
India   &   34,7    &   14,9    &   5   &   0,8 &   0,1 &   0   \\
Indonesia   &   6,8 &   3,3 &   1,4 &   0,3 &   0   &   0   \\
Iran    &   30,2    &   14,2    &   6,2 &   2   &   0,5 &   0,1 \\
Iraq    &   25,3    &   11,5    &   5,1 &   2   &   0,5 &   0   \\
Ireland &   61,2    &   22  &   7   &   1,5 &   0,3 &   0   \\
Isle of Man UK  &   100 &   65,9    &   19,1    &   0   &   0   &   0   \\
Israel  &   90,3    &   78  &   58,6    &   35,7    &   9,4 &   1   \\
Italy   &   99,4    &   91,9    &   58,7    &   19,1    &   1,9 &   0,1 \\
Ivory Coast &   2,5 &   0,9 &   0,3 &   0,1 &   0   &   0   \\
Jamaica &   93,7    &   57,4    &   22,4    &   5,2 &   1   &   0   \\
Japan   &   98,5    &   84,4    &   53,5    &   24  &   5,6 &   1   \\
Jersey UK   &   100 &   100 &   68,4    &   0   &   0   &   0   \\
Jordan  &   27,2    &   15,8    &   9   &   2,1 &   0,4 &   0   \\
Kazakhstan  &   4,3 &   1,9 &   0,8 &   0,2 &   0   &   0   \\
Kenya   &   1,6 &   0,7 &   0,3 &   0,1 &   0   &   0   \\
Kuwait  &   100 &   88,4    &   65  &   39,4    &   13,2    &   3,5 \\
Kyrgyzstan  &   12,4    &   5,3 &   1,5 &   0,1 &   0   &   0   \\
Laos    &   1,7 &   0,7 &   0,3 &   0   &   0   &   0   \\
Latvia  &   34,8    &   13  &   4   &   1,1 &   0,3 &   0   \\
Lebanon &   100 &   66  &   33,3    &   6,7 &   0,3 &   0   \\
Lesotho &   3,4 &   1,2 &   0,4 &   0   &   0   &   0   \\
Liberia &   0   &   0   &   0   &   0   &   0   &   0   \\
Libya   &   8,4 &   4,1 &   1,7 &   0,6 &   0,1 &   0   \\
Liechtenstein   &   100 &   100 &   90,6    &   0   &   0   &   0   \\
Lithuania   &   62,2    &   21,3    &   6,9 &   1,5 &   0,1 &   0   \\
Luxembourg  &   100 &   100 &   100 &   61,2    &   11,6    &   0   \\
Macau P &   100 &   100 &   100 &   100 &   100 &   0   \\
Macedonia   &   56,7    &   19,3    &   5,6 &   0,5 &   0   &   0   \\
Madagascar  &   0,2 &   0,1 &   0   &   0   &   0   &   0   \\
Malawi  &   3,4 &   1,7 &   0,7 &   0,2 &   0,1 &   0   \\
Malaysia    &   22,2    &   11,9    &   5,4 &   1,6 &   0,3 &   0   \\
Mali    &   0,4 &   0,2 &   0,1 &   0   &   0   &   0   \\
Malta   &   100 &   100 &   99,4    &   73,7    &   14,4    &   0   \\
Martinique F    &   100 &   91,9    &   67  &   16,6    &   0   &   0   \\
Mauritania  &   0,2 &   0,1 &   0   &   0   &   0   &   0   \\
Mayotte F   &   7,9 &   3   &   0   &   0   &   0   &   0   \\
Mexico  &   30,5    &   16,1    &   7,2 &   2,4 &   0,6 &   0,1 \\
Moldova &   67,3    &   26,2    &   7,1 &   0,9 &   0,1 &   0   \\
Monaco  &   100 &   100 &   100 &   100 &   63,2    &   0   \\

 \hline
  \end{tabular}
  \end{minipage}
\end{table*}

\begin{table*}
\begin{minipage}{80mm}
%  \centering
  \contcaption{}\label{tab7}
  \begin{tabular}{|lcccccc}
    \hline
        &(1)&(2)&(3)&(4)&(5)&(6)\\
         Country & $\geq 0.11b_{n}$&$\geq 0.33b_{n}$&$\geq b_{n}$&$\geq 3b_{n}$&$\geq 9b_{n}$&$\geq 27b_{n}$\\
         \hline
Mongolia    &   0,3 &   0,1 &   0   &   0   &   0   &   0   \\
Montenegro  &   31,3    &   10,9    &   2,9 &   0,1 &   0   &   0   \\
Montserrat UK   &   56,1    &   15,8    &   0   &   0   &   0   &   0   \\
Morocco &   12,4    &   4,9 &   1,7 &   0,5 &   0,1 &   0   \\
Mozambique  &   0,5 &   0,2 &   0,1 &   0   &   0   &   0   \\
Myanmar &   2,9 &   1,1 &   0,3 &   0,1 &   0   &   0   \\
Namibia &   0,7 &   0,3 &   0,1 &   0   &   0   &   0   \\
Nepal   &   3,1 &   1,2 &   0,3 &   0   &   0   &   0   \\
Netherlands &   100 &   99,1    &   96,7    &   56,8    &   8,3 &   0,6 \\
Netherlands Antilles    &   89,3    &   66,1    &   43,2    &   26,8    &   5,4 &   0   \\
New Caledonia   &   3,2 &   1,3 &   0,6 &   0,3 &   0   &   0   \\
New Zealand &   11,7    &   5   &   2,1 &   0,7 &   0,1 &   0   \\
Nicaragua   &   8,2 &   3,3 &   1,2 &   0,3 &   0   &   0   \\
Niger   &   0,2 &   0,1 &   0   &   0   &   0   &   0   \\
Nigeria &   12,4    &   7,8 &   5,3 &   3   &   1,1 &   0,1 \\
Norfolk Islands AU  &   2,9 &   0   &   0   &   0   &   0   &   0   \\
North Korea &   8,8 &   3,6 &   1,1 &   0,1 &   0   &   0   \\
North. Mariana US   &   0   &   0   &   0   &   0   &   0   &   0   \\
Norway  &   62,8    &   34,6    &   14,6    &   4,4 &   0,9 &   0,1 \\
Oman    &   27,8    &   12,7    &   4,8 &   1,4 &   0,3 &   0   \\
Pakistan    &   30,2    &   19,4    &   7,9 &   0,9 &   0,2 &   0   \\
Panama  &   11,7    &   5,4 &   2,3 &   0,7 &   0,1 &   0   \\
Papua New Guinea    &   2,1 &   1   &   0,5 &   0,2 &   0   &   0   \\
Paraguay    &   4,6 &   2   &   0,9 &   0,4 &   0,1 &   0   \\
Peru    &   3   &   1,3 &   0,5 &   0,2 &   0,1 &   0   \\
Philippines &   12,6    &   6,2 &   2,5 &   0,7 &   0,2 &   0   \\
Poland  &   96,8    &   59,6    &   23,9    &   4,3 &   0,5 &   0   \\
Portugal    &   85,1    &   47,8    &   24,1    &   6,8 &   1,1 &   0,2 \\
Puerto Rico &   99,4    &   98,9    &   97,2    &   68,4    &   11,5    &   2,2 \\
Qatar   &   99,3    &   89,5    &   55,2    &   27,5    &   8,8 &   2,6 \\
Romania &   52,2    &   20,5    &   5,9 &   0,6 &   0   &   0   \\
Russia  &   24,2    &   11,1    &   4,3 &   1,2 &   0,2 &   0   \\
Rwanda  &   2   &   0,8 &   0,3 &   0   &   0   &   0   \\
Saint Kitts e Nevis &   92,3    &   53  &   22,1    &   0   &   0   &   0   \\
Saint Lucia &   60,3    &   29,7    &   10,9    &   0   &   0   &   0   \\
San Marino  &   100 &   100 &   100 &   97,2    &   0   &   0   \\
Saudi Arabia    &   19,3    &   9,8 &   4,5 &   1,7 &   0,4 &   0,1 \\
Senegal &   1,3 &   0,5 &   0,2 &   0,1 &   0   &   0   \\
Serbia  &   72  &   37,3    &   11,8    &   1,1 &   0,1 &   0   \\
Seychelles  &   0   &   0   &   0   &   0   &   0   &   0   \\
Sierra Leone    &   0,6 &   0,2 &   0,1 &   0   &   0   &   0   \\
Singapore   &   100 &   100 &   100 &   100 &   100 &   33,3    \\
Slovakia    &   100 &   98,4    &   64,1    &   4,8 &   0,2 &   0   \\
Slovenia    &   100 &   84,2    &   29,3    &   3,1 &   0,2 &   0   \\
Somalia &   0   &   0   &   0   &   0   &   0   &   0   \\
South Africa    &   13,7    &   6,8 &   3   &   1,2 &   0,3 &   0   \\
South Korea &   99,7    &   98,2    &   79,2    &   32  &   5,9 &   1,2 \\
Spain   &   83,3    &   50,4    &   23  &   7,3 &   1,4 &   0,2 \\
Sri Lanka   &   21,6    &   8,8 &   2,4 &   0   &   0   &   0   \\
St.Vincent-Grenadines   &   29,7    &   13,8    &   2,4 &   0   &   0   &   0   \\
Sudan   &   0,8 &   0,4 &   0,1 &   0   &   0   &   0   \\
Suriname    &   1   &   0,4 &   0,2 &   0   &   0   &   0   \\
Swaziland   &   11  &   4,1 &   1,2 &   0   &   0   &   0   \\
Sweden  &   66,9    &   49,9    &   26,6    &   6,7 &   1,2 &   0,1 \\
Switzerland &   100 &   97,7    &   57,4    &   10,2    &   0,4 &   0   \\
Syria   &   50,3    &   26,6    &   11,2    &   3,9 &   1   &   0,1 \\
Taiwan  &   90,5    &   63,1    &   45,5    &   27,2    &   6,4 &   0,4 \\
Tajikistan  &   13,1    &   5,5 &   1,7 &   0   &   0   &   0   \\
Tanzania    &   1,5 &   0,7 &   0,3 &   0,1 &   0   &   0   \\
Thailand    &   33,6    &   18,1    &   9   &   2,7 &   0,5 &   0,1 \\
Togo    &   1,3 &   0,6 &   0,3 &   0   &   0   &   0   \\

 \hline
  \end{tabular}
  \end{minipage}
\end{table*}

\begin{table*}
\begin{minipage}{80mm}
%  \centering
  \contcaption{}\label{tab8}
  \begin{tabular}{|lcccccc}
    \hline
    &(1)&(2)&(3)&(4)&(5)&(6)\\
          Country & $\geq 0.11b_{n}$&$\geq 0.33b_{n}$&$\geq b_{n}$&$\geq 3b_{n}$&$\geq 9b_{n}$&$\geq 27b_{n}$\\
        \hline
Trinidad and Tobago &   89,8    &   53,7    &   29,6    &   10,7    &   0,1 &   0   \\
Tunisia &   28  &   12,6    &   4,8 &   1,2 &   0,1 &   0   \\
Turkey  &   31,2    &   12,2    &   4,1 &   0,7 &   0,1 &   0   \\
Turkmenistan    &   9,2 &   4,1 &   1,6 &   0,4 &   0,1 &   0   \\
Turks and Caicos Is. UK &   15,6    &   8,7 &   0   &   0   &   0   &   0   \\
Uganda  &   1,1 &   0,5 &   0,2 &   0   &   0   &   0   \\
Ukraine &   62,4    &   31,2    &   11,1    &   1,7 &   0,1 &   0   \\
United Arab Emirates    &   74,6    &   50,9    &   30,2    &   12,8    &   3,1 &   0,7 \\
United Kingdom  &   84,7    &   67,9    &   48,1    &   20,1    &   3,5 &   0,1 \\
United States of America    &   61,8    &   42,7    &   22,5    &   9,2 &   2,6 &   0,6 \\
Uruguay &   14,4    &   5,6 &   2,3 &   0,9 &   0,3 &   0   \\
Uzbekistan  &   23,4    &   13,8    &   6,5 &   1   &   0,1 &   0   \\
Vanuatu &   5,7 &   4,8 &   2,7 &   1,1 &   0   &   0   \\
Venezuela   &   21,6    &   10,8    &   5,1 &   2   &   0,5 &   0   \\
Vietnam &   7   &   2,8 &   1   &   0,3 &   0   &   0   \\
Virgin islands US   &   100 &   93,6    &   77,3    &   53,2    &   0   &   0   \\
Western Sahara  &   0,7 &   0,3 &   0,1 &   0   &   0   &   0   \\
Yemen   &   6,1 &   2,5 &   0,8 &   0,2 &   0   &   0   \\
Zaire   &   0,4 &   0,1 &   0,1 &   0   &   0   &   0   \\
Zambia  &   1,1 &   0,4 &   0,2 &   0   &   0   &   0   \\
Zimbabwe    &   2,6 &   1,1 &   0,5 &   0,1 &   0   &   0   \\
European Union  &   85,3    &   64,8    &   36,7    &   11,5    &   1,7 &   0,1 \\
The World   &       18.7     &  10.9    &   5.3      &  1.8      &  0.4   & 0.1\\
   \hline
  \end{tabular}
  \end{minipage}
\end{table*}

}

%\noindent

\label{lastpage}

\bsp

\end{document}